\documentclass[a4paper,11pt]{article}
\pdfoutput=1 
\usepackage[symbol]{footmisc} 
\usepackage{jheppub} 
	
\usepackage{color, colortbl}
\usepackage[T1]{fontenc}
\usepackage[utf8]{inputenc}
\usepackage[T1]{fontenc}
\usepackage{epsfig,latexsym}
\usepackage{amsmath}
\usepackage[dvipsnames]{xcolor}
\usepackage{adjustbox}
\usepackage{tensor}
\usepackage{graphicx}
\usepackage{verbatim}
\usepackage{mathrsfs}
\usepackage{amssymb}
\usepackage{multirow}
\usepackage{epsfig}
\usepackage{color,colordvi}
\usepackage{appendix}
\usepackage{slashed}
\usepackage{array}
\usepackage{cancel}
\usepackage{float}
\usepackage[font=footnotesize, justification=centering]{caption}
\usepackage{epsf}
\usepackage{amsmath}
\usepackage{tikz-feynhand}
\usepackage{physics}
\usepackage{MnSymbol}
\usepackage{rotating}
\usepackage{multirow}
\usepackage[normalem]{ulem}
\usepackage{tcolorbox}
\usepackage{bbold}
\usepackage{nicefrac}
\DeclareMathAlphabet{\mathpzc}{OT1}{pzc}{m}{it}

\renewcommand{\theequation}{\thesection.\arabic{equation}} \csname
@addtoreset\endcsname{equation}{section}

\newcolumntype{x}[1]{>{\centering\arraybackslash\hspace{0pt}}p{#1}}

\newcommand{\beq}{\begin{equation}}
\newcommand{\eeq}{\end{equation}}
\renewcommand{\[}{\left[}
\renewcommand{\]}{\right]}
\renewcommand{\(}{\left(}
\renewcommand{\)}{\right)}

\newcommand{\be}{\begin{eqnarray}}
\newcommand{\ee}{\end{eqnarray}}
\newcommand{\bea}{\begin{eqnarray}}
\newcommand{\eea}{\end{eqnarray}}
\newcommand{\bi}{\begin{itemize}}
\newcommand{\ei}{\end{itemize}}
\newcommand{\ben}{\begin{enumerate}}
\newcommand{\een}{\end{enumerate}}
\def\bes{\begin{equation*}}
\def\ees{\end{equation*}}
\def\bead{\begin{aligned}}
\def\eead{\end{aligned}}
\def\bmat{\left(\begin{matrix}}
\def\emat{\end{matrix}\right)}

\def\Re{\text{Re}}

\def\cA{{\cal A}}

\def\cL{{\cal L}}

\def\beq{\begin{equation}}
\def\eeq{\end{equation}}
\def\bes{\begin{equation*}}
\def\ees{\end{equation*}}
\def\bead{\begin{aligned}}
\def\eead{\end{aligned}}
\def\bmat{\left(\begin{matrix}}
\def\emat{\end{matrix}\right)}
\def\cA{{\cal A}}
\def\cL{{\cal L}}

\renewcommand{\[}{\left[}
\renewcommand{\]}{\right]}
\renewcommand{\(}{\left(}
\renewcommand{\)}{\right)}

\title{Causality constraints on nonlinear supersymmetry}

\author[a]{Quentin Bonnefoy,}
\author[b]{Gabriele Casagrande}
\author[b,c]{and Emilian Dudas}
\affiliation[a]{Deutsches Elektronen-Synchrotron DESY, Notkestr. 85, 22607 Hamburg, Germany}
\affiliation[b]{Centre de Physique Th{\'e}orique, {\'E}cole Polytechnique, CNRS and IP Paris, 91128 Palaiseau Cedex, France 
}
\affiliation[c]{{CERN, Theory Division, Geneva, Switzerland}}

\emailAdd{quentin.bonnefoy@desy.de, gabriele.casagrande@polytechnique.edu,emilian.dudas@polytechnique.edu}

\abstract{It is well-known that gravitino propagation in standard supergravities is free of any causality problems. However, two issues related to gravitino propagation were recently uncovered in specific supergravities with nonlinear supersymmetry. One of them concerns potential acausality/superluminality, whereas the second one arises from the vanishing of the sound speed at specific points during inflation. The former is famously related to positivity constraints on specific EFT operators, derived from dispersion relations on the energy-growing part of scattering amplitudes, and indeed we show that subluminality constraints for the gravitino are related via the equivalence theorem to positivity bounds in low-energy goldstino actions. However, the former are stronger, in the sense that they apply to functions of the scalar fields not only in the ground state, but for any field values such as those scanned by time-dependent solutions, unlike bounds derived from $2\to 2$ scattering amplitudes in the vacuum. We also argue that nontrivial causality constraints arise only in the case where nonlinear supersymmetry in the matter sector is encoded into superfield constraints which do not seem to arise from microscopic two-derivative lagrangians, in particular for the orthogonal constraint used to build minimal models of inflation in supergravity. This allows us to propose simple alternatives which maintain the minimality of the spectra and are causal in all points of the theory parameter space. We also discuss minimal supergravity models of inflation along these lines.}

\begin{document} 
\begin{flushright}
DESY-22-103\\
CPHT-RR040.062022\\
CERN-TH-2022-096
\end{flushright}

\maketitle

\flushbottom

\section{Introduction}\label{section:intro}

The consistency of low-energy actions for spin-3/2 Rarita-Schwinger fermions has a long and interesting history. It was shown in the sixties that a charged spin 3/2 field in an electromagnetic background has acausal propagation features \cite{Velo:1969bt}. When supergravity (SUGRA) was invented in 1976 \cite{Freedman:1976xh}, this issue was immediately raised and the minimal supergravity was shown to be causal \cite{Deser:1977uq} due to its specific interactions. Much later, it was shown that a charged gravitino propagation is also causal in gauged supergravities \cite{Deser:2001dt}, solving the original Velo-Zwanziger problem.  

When supersymmetry is realized nonlinearly, causality of gravitino propagation has to be re-examined. Starting from a standard two-derivative SUGRA and taking the scale of supersymmetry breaking to large values, one expects to decouple superpartners without generating any causality problems. If one instead write the most general nonlinearly-supersymmetric lagrangians, there is a priori no guarantee of their microcopic origin and causality problems can arise. This does not occur for the simplest (and oldest) Volkov-Akulov realization \cite{Volkov:1973ix}. In this case, supersymmetry is realized 
as a diffeomorphism, starting initially with a two-derivative action coupled to a metric/vierbein. The goldstino/gravitino couplings arise through an appropriately defined goldstino-dependent vierbein, which in the end induce also higher-derivative operators containing the goldstino.       

A particularly simple formalism to build  nonlinear supersymmetry (SUSY) or SUGRA lagrangians is by using superfield \textit{constraints} \cite{Rocek:1978nb, Lindstrom:1979kq,Komargodski:2009rz,DallAgata:2015zxp}: this procedure allows to fix/remove some of the superfields components, simulating the decoupling of a massive degree of freedom while allowing one to use the standard supersymmetric lagrangians. Constraints appropriate to remove scalar, fermions and even auxiliary fields were proposed. Nonlinear supergravities along these lines were soon constructed afterwards 
\cite{Dudas:2015eha,Bergshoeff:2015tra,Bandos:2015xnf,Bandos:2016xyu}. Such setups were widely used to construct more minimal inflationary models in supergravity \cite{nonlinear-sugra}, in particular models with a minimal particle spectrum: a graviton, a massive gravitino and a (real scalar) inflaton  \cite{Kallosh:2014via,Ferrara:2015tyn,Carrasco:2015iij}, using the so-called orthogonal constraint. It was subsequently realized that such models have peculiar features, like the possibility that the gravitino sound speed becomes zero and generates an unbounded production of gravitinos \cite{Hasegawa:2017hgd,Kolb:2021xfn,Dudas:2021njv,Terada:2021rtp,Antoniadis:2021jtg} and even superluminal propagation \cite{Dudas:2021njv}, both arising for the longitudinal component of the gravitino\footnote{The possibility of the gravitino to propagate "slowly" was noticed in \cite{Benakli:2014bpa}.}.   

In this paper we shed more light on the sound speed of gravitino in such constructions (assuming the presence of a nilpotent superfield, which parametrizes supersymmetry breaking throughout our paper). First, by using the equivalence theorem between the longitudinal component of the gravitino and the goldstino, valid at energies $E \gg m_{3/2}$ \cite{Fayet:1986zc} , we write down low-energy goldstino lagrangians coupled to scalars, eventually to be identified with the inflaton in supergravity.  Our results apply however to gravitino couplings to any scalars, not only the inflaton. 
The superfield constraints generate, in the low-energy theory with broken SUSY, higher-derivative operators which are subject to positivity bounds \cite{Adams:2006sv, Bellazzini:2016xrt} in order for the theory to respect causality, analyticity and crossing-symmetry (see also \cite{Bellazzini:2014waa,Trott:2020ebl,Caron-Huot:2020cmc,Alberte:2020jsk,Bellazzini:2015cra,Bellazzini:2020cot,Wang:2020jxr,deRham:2017avq}). Positivity and superluminality are well-known to be connected \cite{Adams:2006sv}, and we confirm that the positivity constraints map precisely into subluminality constraints derived directly in SUGRA from the gravitino sound speed, in the limit of decoupling gravity $M_P \to \infty$. However, whereas positivity constraints are valid only in the ground state, subluminality of sound speed in supergravity should be respected throughout the time-dependent scalar field evolution, which is a priori a  stronger constraint. Knowing that a microscopic two-derivative SUGRA lagrangian has no acausal propagation problems, we attempt to identify the origin of the subluminality constraints as a potential unphysical feature of the orthogonal constraint. Our understanding is that, whereas the "removal" of a scalar or fermion can be easily understood microscopically as standard decoupling of heavy particles, removal of an auxiliary field requires higher-derivative couplings directly in the UV, as proposed already in \cite{DallAgata:2016syy}. We substantiate this claim in two ways. First, by imposing alternative constraints that remove a real scalar and its fermionic partner only (not the auxiliary field) and by constructing low-energy goldstino lagrangians coupled to the real scalar: in this improved framework we find no non-trivial causality condition to be imposed on the low-energy theory. Alternatively, we propose a procedure to integrate-out an auxiliary field at the superfield level, which yields an effective action of superfields constrained so as to not contain auxiliary fields anymore. We show explicitly that higher-derivative terms are generated in the action by that procedure, and that they precisely cancel the "wrong-sign" terms in the operators subject to positivity bounds. The result leads to an effective action which trivially satisfies the (twice-subtracted and forward) positivity bounds for all values of fields and couplings. Finally, we rely on these insights to construct simple realistic minimal models of inflation in SUGRA with causality  respected everywhere in the theory parameter space. Let us nevertheless stress that we are not claiming that any two-derivative SUGRA theory supplemented with the orthogonal constraint cannot be realized in a microscopic theory. If the causality constraints that we discuss later on are satisfied {\it identically}, i.e. for any values of the parameters of the theory, this is perfectly acceptable. For this to happen, the functions defining the effective action (superpotential and K\"ahler potential) should have some restricted form. This is a priori possible to obtain from a microscopic theory. Our precise claim is that any non-trivial causality constraints on the theory parameters should not come from a microscopic theory, or in a modern language, is in the swampland \cite{Vafa:2005ui}. Our alternative constructions have the particular feature that for {\it the most general}  superpotential and K\"ahler potentials, the theory is causal for arbitrary values of the parameters of the theory. 

The paper is organized as follows. In section \ref{section:equivalence} we review the gravitino equivalence theorem and the gravitino sound speed in supergravity. We compute the sound speed in various examples with nonlinear supersymmetry and identify cases with potential superluminal propagation.
Section \ref{section:goldstino-actions} constructs low-energy goldstino actions coupled to matter, with nonlinearly-realized supersymmetry generated by constrained superfields. We
use these results to work out causality/positivity constraints in various cases, which turn out to reproduce the corresponding SUGRA subluminality constraints on the gravitino sound speed in the decoupling limit $M_P \to \infty$, and in our main example, the orthogonal constraint,  even exactly.  
In Sections \ref{section:twoconstraints} and 
\ref{section:orthogonal-nocondition} we propose two different, but equivalent, ways to obtain goldstino actions from microscopic theories. The one in  Section \ref{section:twoconstraints} starts from a two-derivative low-energy action and, instead of using the orthogonal constraint which eliminates a (imaginary) scalar, its fermionic partner and the auxiliary field, uses the alternative constraints which eliminate only the scalar and the fermion \cite{DallAgata:2016syy} . The second approach put forward in
Section \ref{section:orthogonal-nocondition} starts from a microscopic two-derivative action and uses the presence of nonlinear SUSY to integrate the auxiliary field out of a chiral multiplet in a superspace/superfield fashion, instead of the standard component field algebraic procedure. The procedure leads to an action of superfields constrained so that their auxiliary fields are eliminated. In particular, this is able to generate the orthogonal constraint, but at the same time it generates higher-derivative terms which remove all potentially non-trivial causality conditions, yielding a completely safe action. We believe that the two procedures are equivalent, as they assume the same microscopic action and superfield content and subsequently only amount to solving equations of motion.  Whereas the results apply beyond models of slow-roll inflation, we use them to construct simple examples of completely causal minimal SUGRA models of inflation in Section \ref{section:inflation}. We summarize our findings and present our perspectives in
the Conclusions, Section \ref{section:conclusions}. Appendix \ref{appendix:solutionsConstraints} displays the complete solution of the alternative constraints put forward in Section \ref{section:twoconstraints}, while Appendix \ref{appendix:IRlagNoF} presents the derivation of the main result of Section \ref{Fint}.

\section{Gravitino sound speed, goldstino and the equivalence theorem}\label{section:equivalence}

The gravitino, spin $3/2$ superpartner of the graviton\footnote{We are only considering supersymmetry/SUGRA models with minimal supersymmetry in four dimensions.}, has four degrees of freedom if supersymmetry is broken. Analogously to the Higgs mechanism, there is a super-Higgs mechanism in which a massless gravitino with helicities $\pm 3/2$ absorbs the goldstino spin $1/2$ fermion related to supersymmetry breaking in order to form a massive gravitino with four helicity states. The states  of helicity $\pm 3/2$ form the transverse gravitino, whereas the helicity $\pm 1/2$ states are the longitudinal components. By representing the spin $3/2$ in a basis of  a spin $1$ and a spin $1/2$ states, one can write
\begin{eqnarray}
&& | \frac{3}{2} , \pm \frac{3}{2} \rangle = 
| 1 , \pm 1 \rangle \otimes 
| \frac{1}{2} , \pm \frac{1}{2} \rangle \ , \nonumber \\
&& | \frac{3}{2} , \pm \frac{1}{2} \rangle = 
\sqrt{\frac{2}{3}} \ | 1 ,  0 \rangle \otimes 
| \frac{1}{2},  \pm \frac{1}{2} \rangle + 
\sqrt{\frac{1}{3}}
\ | 1 , \pm 1 \rangle \otimes 
| \frac{1}{2} , \mp \frac{1}{2} \rangle \ . \label{eq:equiv1}
\end{eqnarray}
The gravitino $\Psi_\mu$ couples in SUGRA to matter fields via the supercurrent
\begin{equation}
\frac{1}{M_P} \Psi^\mu J_\mu  \ . \label{eq:equiv2}   
\end{equation}
Whereas the transverse gravitino couples with gravitational strength, the longitudinal component couples stronger. Indeed, the gravitino equivalence theorem \cite{Fayet:1986zc} states that at energies $E \gg m_{3/2}$ the amplitudes for the $\pm 1/2$ helicities can be computed by using the goldstino $G$ via the substitution
\begin{equation}
\Psi^\mu \to \frac{1}{m_{3/2}} \partial^\mu G  \ . \label{eq:equiv3}  
\end{equation}
The longitudinal gravitino component couplings can then be replaced by
\begin{equation}
 \frac{1}{f} \partial^\mu G \ J_\mu = - \frac{1}{f} G \  \partial^\mu J_\mu   \ , \label{eq:equiv4}
\end{equation}
where $f = m_{3/2} M_P$ is the supersymmetry breaking scale. The couplings of the longitudinal component are therefore enhanced compared to the transverse component, which are gravitationally suppressed. One can take in particular the decoupling limit of gravity $M_P \to \infty$ with fixed $f$. 
On-shell, $\partial^\mu J_\mu$ is proportional to soft terms, hence goldstino couplings to matter can also be written as being proportional
to $m_\textup{soft}/f$. When superpartners are very heavy, their virtual exchange generate effective operators coupling the goldstino and the light matter fields. In the original Volkov-Akulov (VA) formulation they are determined geometrically, typically of dimension eight   and of the type
\begin{equation}
L_{VA} = \frac{i}{2f^2}(G\sigma^\mu\partial^\nu\bar{G} - \partial^\nu G\sigma^\mu\bar{G}) \ T_{\mu\nu} \ , \label{eq:equiv5}
\end{equation}
where $G$ is the goldstino field and $T_{\mu\nu}$ is the energy momentum tensor of the light matter fields with heavy superpartners. Whereas dimension-eight operators of this type are generally subject to positivity conditions coming from general arguments of unitarity and causality, the ones generated via the VA prescription, produced starting from a two-derivative action coupling to the VA  vierbein, come automatically with the correct signs. The same will be true for effective operators obtained by tree-level exchange of heavy superpartners starting from the supercurrent couplings (\ref{eq:equiv4}).  For other nonlinear supersymmetry realizations, in particular by constrained superfields, the positivity/causality constraints  can turn out to be non-trivial. Through the equivalence theorem, such positivity conditions are manifest in the corresponding SUGRA lagrangians, in particular in the sound speed of the gravitino. 

It was shown in \cite{Kolb:2021xfn} that the transverse gravitino in SUGRA propagates at the speed of light. On the other hand, the longitudinal component has a more complicated sound speed $c_s$, which can be written in general in terms of the energy density $\rho$ and pressure $p$, according to the formula
\begin{equation}
c_s^2 = \frac{(p-3 m_{3/2}^2)^2}{(\rho+3 m_{3/2}^2)^2} + \frac{4 {\dot m_{3/2}}^2}{(\rho+3 m_{3/2}^2)^2} \ . \label{eq:cs_1}  
\end{equation}
For standard SUGRA theories with two-derivative couplings, it was shown in full generality in \cite{Dudas:2021njv} that $c_s \leq 1$. For completeness we review the argument here. 
Ref.~\cite{Kolb:2021xfn} provides a useful expression for  $c_s$ in any ${\cal N}=1$ supergravity model with chiral superfields, \begin{equation} 
c_s^2 = 1 - \frac{4}{
\left(\vert\dot{\varphi}\vert^2 + \vert F \vert^2 \right)^2} \, 
\left\{ 
 |\dot{\varphi}|^2  |F|^2  - 
\left\vert \dot{\varphi}  \cdot F^*  \right\vert^2 \right\}  \;, 
\label{vs2}
\end{equation} 
where $\varphi$ and $F$ correspond to vectors of scalar\footnote{Formula (\ref{vs2}) actually holds for real scalar fields $\varphi$. However, the formula that we use to compute the gravitino sound speed in the various models analyzed throughout the paper is the more general (\ref{eq:cs_1}).} and auxiliary fields respectively, and the $F$-terms are given by  
\begin{equation}
F^i \equiv - {\rm e}^{K/2} K^{ij^*} \, D_{j^*} W^* \;, 
\label{DWF}
\end{equation}
where in a standard supergravity notation, $K^{ij^*}$ is the inverse of the K\"ahler metric 
\begin{equation}
K_{ij^*} \equiv \frac{\partial^2 K}{\partial \varphi^i \, \partial \varphi^{j*}} \;,
\label{Kij}
\end{equation} 
while 
\begin{equation}
D_i W \equiv \frac{\partial W}{\partial \varphi^i} +  \frac{\partial K}{\partial \varphi^i} \, W \,. 
\end{equation} 
Throughout our paper, we will consider the most general superpotential $W$ and K\"ahler potential $K$, not necessarily renormalizable. The dot operator in \eqref{vs2} denotes a  scalar product with the K\"ahler metric (\ref{Kij}), namely $\vert \dot{\varphi} \vert^2 = \dot{\varphi}^i \, K_{ij^*} \, \dot{\varphi}^{j*}$, and similarly for the other terms\footnote{\label{MPfootnote}We work in Planck units $M_P=1$, but restoring $M_P$ whenever needed can simply be achieved by dimensional analysis.}.
Finally, due to the Cauchy-Schwarz type inequality
$|\dot{\varphi}|^2  |F|^2  \geq 
\left\vert \dot{\varphi}  \cdot F^*  \right\vert^2$, causality $c_s \leq 1$ is always guaranteed to hold. The argument can be extended to include D-term contributions to the scalar potential, which change the sound speed according to
 \begin{equation} 
c_s^2 = 1 - \frac{4}{
\left(\vert\dot{\varphi}\vert^2 + \vert F \vert^2  + 
\frac{1}{2} D^2 \right)^2}\, 
\left\{ 
 |\dot{\varphi}|^2  \(|F|^2 + \frac{1}{2} D^2\) - 
\left\vert \dot{\varphi}  \cdot F^*  \right\vert^2 \right\} 
\; .
\label{vs02}
\end{equation}    
The subluminality condition is strongest when $D=0$ and therefore models with only F-terms are the most constraining.

The proof above does not apply to higher-derivative theories and for some realizations with nonlinear supersymmetry, which we will analyze here from several perspectives. 
The pressure and energy density associated to the scalar $\varphi$ are 
given by
\begin{equation}
p = K_{\Phi \bar \Phi} {\dot \varphi}^2 - V(\varphi) \quad , \quad  \rho = K_{\Phi \bar \Phi} {\dot \varphi}^2 + V(\varphi)
\ . \label{eq:cs2}
\end{equation}
It was shown in \cite{Dudas:2021njv} that the sound speed 
(\ref{eq:cs_1}) applies actually to the case where there is no fermion/inflatino in the spectrum. A particularly interesting example in this class of models is the {\it orthogonal constraint} defined by
\begin{equation}
S (\Phi - {\bar \Phi}) = 0 \ , \label{eq:orthog1}
\end{equation}
where $\Phi$ is a chiral superfield and $S$ is a chiral nilpotent goldstino superfield, whose scalar component $s$ is expressed in terms of a goldstino bilinear. This constraint turns the imaginary part of the scalar, the fermion and
the auxiliary field contained in $\Phi$ into functions of the goldstino $G$ and the real part $\varphi$ of the scalar in $\Phi$, which vanishes when $G\to 0$ (see Section \ref{section:goldstino-actions} for more details). Decomposing $\Phi = \mathcal{A} + i\mathcal{B}$ ($\varphi$ is then the lowest, scalar component of $\mathcal{A}$ and $\cal B$ vanishes when $G\to 0$), the generic SUGRA lagrangian is defined by \cite{Ferrara:2015tyn}
\begin{equation}
K = h\(\mathcal{A}\)\mathcal{B}^2 + S {\bar S} \quad , \quad
W = f (\Phi) S + g(\Phi) \ , \label{eq:orthog2}
\end{equation} 
where $h$ is a real function, while $f,g$ are holomorphic. As said above, the sgoldstino scalar $s$ and the auxiliary field $F_{\phi}$ in $\Phi$ are expressed in terms of fermionic terms. This implies in particular that only the auxiliary field of $S$ contributes to the scalar potential, which is given by
\begin{equation}
V = |f (\varphi)|^2 - 3 |g (\varphi)|^2 \ , 
\label{eq:orthog3}
\end{equation}

\noindent and the sound speed (\ref{eq:cs_1}) reads\footnote{We stress here that to compute the sound speed in this model, we need to make use of the formula (\ref{eq:cs_1}), and not of (\ref{vs2}): while the former applies in general, in particular for nonlinear realizations of supersymmetry such as (\ref{eq:orthog2}), the latter applies only for linearly-realised SUSY.}
\begin{equation}
c_s^2 = 1 - \frac{4 {\dot \varphi}^2}{\(\frac{h (\varphi)}{2} {\dot \varphi}^2 + |f (\varphi)|^2\)^2} \(\frac{h (\varphi)}{2} |f(\varphi)|^2 - |g'(\varphi)|^2\) \ .  \label{eq:orthog4}
\end{equation}
Causality imposes therefore a nontrivial condition,
\begin{equation}
\frac{h(\varphi)}{2} |f(\varphi)|^2 \geq |g'(\varphi)|^2  \ .  \label{eq:orthog5}   
\end{equation}
The causality condition (\ref{eq:orthog5}) should be respected not only in the ground state, but on any time-dependent solution for the scalar $\varphi$, being them relevant for inflation or not. Therefore the value of $\varphi$ in (\ref{eq:orthog5}) refers to all possible scalar field values attained during the time evolution. Notice also that the causality/subluminality condition
(\ref{eq:orthog5}) has to be imposed
even for arbitrarily small velocities $\dot \varphi$. We stress that all mass scales in $W,K$ are implicitly assumed to be independent on gravity, i.e. on the Planck mass $M_P$. As such, $W$ and $K$ retain the same form in the limit of decoupling gravity $M_P \to \infty$ with fixed $f$, which defines the goldstino lagrangians upon using the equivalence theorem\footnote{The study of goldstino lagrangians and sound speed for the orthogonal constraint from the viewpoint of the equivalence theorem was already studied in \cite{Kahn:2015mla} in the particular case of a constant gravitino mass.}. This implies in particular that the causality condition (\ref{eq:orthog5}) should remain the same after decoupling gravity, since $M_P$ does not appear anywhere in it. The same is actually true also for the orthogonal constraint gravitino/goldstino sound speed (\ref{eq:cs_1}).

It should be clear from our discussion that the issue is not really tied to inflationary models and is more general. For example, one could consider models with complex scalars, defined by the constraint $S \bar {H} = {\rm chiral}$, where $H$ is another chiral superfield. This constraint eliminates its fermion and auxiliary field, leaving a light complex scalar. 
In this case, the most general model is characterised by the following K\"ahler potential and superpotential,
\beq
    K=\xi(H,\bar{H})S\bar{S}+\kappa(H,\bar{H}) \ , \quad W=f(H)S+g(H) \ , \label{eqn:cs_superpot}
\eeq
where $\xi,\kappa$ are real functions, while $f,g$ are holomorphic ones. Using again (\ref{eq:cs_1}), we see that subluminality imposes now the following constraint on the theory parameter space,
\begin{equation}
\kappa_{H \bar H} |\dot H|^2  |f|^2 
\geq \xi \ |{\dot H} (g' + \frac{g}{2} \kappa_{H}) + {\dot {\bar H}}  \frac{g}{2} \kappa_{{\bar H}}|^2 \ , \label{eqn:cs_sublu}
\end{equation}
where $\kappa_{H} = \partial_H \kappa$, etc. Any nontrivial time dynamics for the scalar $H$ will lead to a nontrivial causality constraint. Differently from (\ref{eq:orthog5}) however, in (\ref{eqn:cs_sublu}) the terms proportional to $ \kappa_{H},\kappa_{\bar H}$ arise from the SUGRA lagrangian with a factor of $1/M_P^2$, which we kept implicit as it can be reinstated on dimensional grounds (see footnote \ref{MPfootnote}).


The presence of a nontrivial causality condition suggests that the microscopic origin of the orthogonal constraint (\ref{eq:orthog1}) is not a standard two-derivative theory. As decoupling a heavy scalar and heavy fermion should not be difficult to obtain, it becomes intuitively clear that the problem comes from the "removal" of the auxiliary field via the constraint. This claim can be tested by imposing constraints that remove only the scalar and the fermion, but not the auxiliary field. As shown in \cite{DallAgata:2016syy}, this can be achieved by imposing the constraints
\begin{equation}
S {\bar S}   (\Phi - {\bar \Phi}) = 0 \ , \quad\quad 
S {\bar S}   D_{\alpha}\Phi  = 0 \ ,  \label{eq:orthog6}
\end{equation}
where the first constraint in (\ref{eq:orthog6}) removes the imaginary part of the scalar and the second one removes the fermion in $\Phi$. Whereas the physical spectrum of such a model is the same (minimal) one as that obtained with the orthogonal constraint, the lagrangian is different. In particular, both auxiliary fields are now determined in the usual SUSY algebraic way, so that the scalar potential reads
\begin{equation}
V = |f(\varphi)|^2 + \frac{2}{h(\varphi)} |g'(\varphi)|^2 - 3 |g(\varphi)|^2 \ ,  
\label{eq:orthog7}    
\end{equation}
and the sound speed changes accordingly into
\begin{equation}
    c_s^2 = 1 - \frac{2 h(\varphi) {\dot \varphi}^2 |f(\varphi)|^2 }  {\(\frac{h(\varphi)}{2} {\dot \varphi}^2 + |f(\varphi)|^2 + \frac{2}{h(\varphi)} |g'(\varphi)|^2\)^2} \ .
    \label{eq:orthog8}    
\end{equation}
which is always subluminal $c_s \leq 1$ since $h$ is positive definite.  In this case causality is therefore automatically satisfied. We interpret this as evidence that the superfield constraints (\ref{eq:orthog6}) can arise from a two-derivative microscopic lagrangian with linear SUSY. On the other hand, in the case of the orthogonal constraint, the nontrivial causality condition (\ref{eq:orthog5}) probably implies that the original high-energy lagrangian should be supplemented with higher-derivative terms which ensure causality for all points in the theory parameter space.  

Comparing subluminality conditions in SUGRA with low-energy goldstino causality conditions, as well as providing evidence for the claims above, are the main subjects for the rest of our paper. Before embarking into more technical considerations on the goldstino lagrangians, we repeat that we expect a causality condition of the type (\ref{eq:orthog5}) to be captured exactly by goldstino actions via the equivalence theorem, since it is independent of $M_P$. On the other hand, for a SUGRA causality condition of the type   (\ref{eqn:cs_sublu}), which depends explicitly on $M_P$, we expect that causality arguments from the goldstino action only agree in the $M_P\to \infty$ limit. Turning the argument around, by imposing causality in SUGRA, we find a small (i.e., order $1/M_P^2$) seemingly violation of causality in the goldstino actions. However, apparent gravitational violations of causality are well-studied and were argued not to be necessarily inconsistent (see for example \cite{Alberte:2020jsk}).       

In addition, when there is no inflatino in the spectrum,  there is the possibility that the gravitino sound speed becomes zero and generates an unbounded production of gravitinos \cite{Hasegawa:2017hgd,Kolb:2021xfn,Dudas:2021njv,Terada:2021rtp,Antoniadis:2021jtg}. This is the case both for the orthogonal constraint (\ref{eq:orthog1}) and for the alternative constraints (\ref{eq:orthog6}). In both cases, the sound speed becomes zero at particular points of the inflationary trajectory which satisfy
\begin{equation}
g'(\varphi) = 0 \quad , \quad  \frac{h(\varphi)}{2} {\dot \varphi}^2 = |f(\varphi)|^2 \ .  \label{eq:zerocs}   
\end{equation}
In the case of the orthogonal constraint, the complete sound speed (\ref{eq:orthog4}) is independent of $M_P$, hence the exact formula can be captured exactly from the low-energy goldstino lagrangians using the equivalence theorem, which we explicitly confirm below. Therefore, goldstino lagrangians in this case capture exactly both the causality/subluminality condition and the vanishing of the sound speed of gravitino in supergravity. For cases where the sound speed in SUGRA contains $M_P$ explicitly, goldstino lagrangians capture only the limit $M_P \to \infty$ of the different formulae.

\section{Low-energy goldstino lagrangians}\label{section:goldstino-actions}

As we anticipated above, all the models we are going to study involve one particular chiral superfield, subject to the \textit{nilpotency} condition \cite{Komargodski:2009rz}
\begin{equation}
	S^2=0 \ .
\label{eqn:const_nil}
\end{equation}
The solution of (\ref{eqn:const_nil}) is
\begin{equation}
	S=\frac{G^2}{2F_S}+\sqrt{2}\theta\,G+\theta^2 F_S \ ,
\label{eqn:goldstino_sf}
\end{equation}
i.e. the scalar field component $s$ of $S$ is not independent anymore, but is a bilinear in the goldstino. $S$ is the field responsible for spontaneous supersymmetry breaking, with the fermion $G$ being identified as the goldstino. SUSY is actually realized nonlinearly \cite{Wess:1992cp} and it has been shown in \cite{Komargodski:2009rz} that writing the most general two-derivative theory for $S$ and subsequently imposing the constraint leads to a goldstino lagrangian which is equivalent to the original Volkov-Akulov theory  \cite{Volkov:1973ix}. The latter is perfectly causal, while the nilpotent constraint on $S$ does not constrain its auxiliary field, which further supports our claim linking the loss of causality to the removal of auxiliary fields by superfield constraints\footnote{The fact that the constraint $S^2=0$, equivalent to the VA dynamics, does not lead to any positivity problem follows from the fact that it is known to arise from UV models, upon integrating the heavy scalar partner of $G$ \cite{Komargodski:2009rz}. Such constraints are genuinely equivalent to constraint of the form $\abs{\Phi}^2=v^2$ for a multiplet of scalar fields $\Phi$, which arise in the mapping from linear to nonlinear sigma models.}.

All the models we are going to consider will be based on the goldstino superfield (\ref{eqn:goldstino_sf}), which not only breaks SUSY but also allows to define other constraints for different superfields, decoupling some field components and realizing SUSY nonlinearly in the corresponding multiplets. Let us outline the general procedure to build such models. Since we will consider only chiral superfields of components $(\phi^i, \chi^i,F^i)$, the starting lagrangian is \cite{Wess:1992cp, Zumino:1979et}
\begin{equation}
\begin{aligned}
    \mathcal{L}=&K_{i\bar{j}}\left[F^i \bar{F}^{\bar{j}}+\partial_\mu\phi^i\partial^\mu\bar{\phi}^{\bar{j}}+\left(\frac{i}{2}\partial_\mu\chi^i\sigma^\mu\bar{\chi}^{\bar{j}}+h.c.\right)\right]+\left[W_i F^i-\frac{1}{2}W_{ij}\chi^i\chi^j+h.c.\right]\\
    &+\left[\frac{i}{4}K_{ij\bar{k}}\left(\chi^i\sigma^\mu\bar{\chi}^{\bar{k}}\partial_\mu\phi^j+\chi^j\sigma^\mu\bar{\chi}^{\bar{k}}\partial_\mu\phi^i+2i\chi^i\chi^j\bar{F}^{\bar{k}}\right)+h.c.\right]+\frac{1}{4}K_{ij\bar{k}\bar{l}}\,\,\chi^i\chi^j\bar{\chi}^{\bar{k}}\bar{\chi}^{\bar{l}},
\end{aligned}
\label{eqn:L_general}
\end{equation}
where $K$ and $W$ are, respectively, the K\"ahler potential and the superpotential, and the subscripts denote their derivatives with respect to the chiral fields. Given a set of superfield constraints, the first step is then to insert their solution in (\ref{eqn:L_general}). The second step consists in integrating out the auxiliary fields, which is non-trivial here because the dependence of the lagrangian (\ref{eqn:L_general}) on the auxiliary fields, and especially on that of the goldstino (see (\ref{eqn:goldstino_sf})), is not polynomial anymore. One typically searches for a solution of the auxiliary field equations of motion (EoM) as a (finite) series of operators with increasing numbers of fermions (see \cite{Komargodski:2009rz, Bergshoeff:2015tra, Kallosh:2015tea} for examples). The resulting lagrangian only depends on the dynamical fields.

In what follows, we will restrict ourselves to two kinds of truncations of the lagrangians. When studying the goldstino sound speed, we only consider operators quadratic in the goldstino field (which we refer to as the "quadratic limit" throughout the discussion), which are those driving the free propagation of the goldstino on a generic scalar background. When studying instead the positivity bounds on $2\to 2$ scattering amplitudes, we can focus on operators involving at most four fields (called the "four-fields limit"). We therefore find it convenient to consider two operator bases, related by field redefinitions: the one which naturally arises from the procedure outlined in the previous paragraph and generates all (finitely many) terms quadratic in the goldstino field, and another one where most three-point couplings between goldstini and scalars have been traded for four-point couplings. The former is more adapted to a quick extraction of the goldstino sound speed, while in the latter  one can directly read off from the associated Wilson coefficients the combination of coefficients constrained by positivity bounds. In any case, the results are physical and can be obtained in any operator basis.  

\subsection{Light real scalar: the orthogonal constraint}
\label{section:orthogonal}

The first model we discuss involves the goldstino superfield (\ref{eqn:goldstino_sf}) coupled to a chiral superfield $\Phi$ subject to what is called the \textit{orthogonal constraint},
\begin{equation}
	S\left(\Phi-\bar{\Phi}\right)=0 \ .
\label{eqn:o_const}    
\end{equation}
This constraint, which implies also $\left(\Phi-\bar{\Phi}\right)^3=0$, fixes the components of $\Phi$, calling the scalar field $\phi\equiv A+iB$, to be \cite{Komargodski:2009rz}
\beq
    B=\frac{G\sigma^\mu\bar G} {2\abs{F_S}^2}\partial_\mu A\ , \quad \chi^\phi=i\sigma^\mu\frac{\bar G}{\bar{F}_S}\partial_\mu\phi\ , \quad F^\phi=-\partial_\nu\left(\frac{\bar G}{\bar{F}_S}\right)\bar{\sigma}^\mu\sigma^\nu \frac{\bar G} {\bar{F}_S} \partial_\mu A-\frac{1}{2}\left(\frac{\bar G}{\bar{F}_S}\right)^2\Box A \ . 
\label{eqn:orthogonalSolution}
\eeq
The resulting spectrum involves only the goldstino $G$ and a real scalar $A$: for this reason, models with such constrained superfields have been largely studied from a cosmological viewpoint as minimal models of inflation, with $A$ being identified with the inflaton (see for instance \cite{Ferrara:2014kva, Ferrara:2015tyn, Carrasco:2015iij, Kolb:2021xfn, Dudas:2021njv}).

The most general theory involving such constrained superfields was given in \eqref{eq:orthog2}, i.e. it is completely characterized by a real function $h(\Phi)$ and two holomorphic ones, $f(\Phi)$ and $g(\Phi)$\footnote{The precise form of the functions $h$, $f$ and $g$ and their combinations is left implicit in what follows.}. We first notice that because of the expression of the K\"ahler potential, the second line of (\ref{eqn:L_general}) does not give any contribution, neither in the quadratic nor in the four-field limit. The starting lagrangian turns out to be\footnote{We follow the conventions of \cite{Bilal:2001nv} in what follows.}
\begin{equation}
    \begin{aligned}
    \mathcal{L}=&\, |F_S|^2+\frac{h}{2}\partial_\mu A\partial^\mu A+\left(\frac{i}{2}\partial_\mu G \sigma^\mu\bar{G}+h.c.\right)-\frac{1}{4}\frac{{\bar G}^2}{\bar{F}_S}\Box\left(\frac{G^2}{F_S}\right)\\
     &+\frac{h}{2}\left[i\, \partial^\mu\left(\frac{G}{F_S}\right)\sigma^\nu\left(\frac{\bar{G}}{\bar{F}_S}\right)\partial_\mu A\partial_\nu A-\frac{i}{2}\partial_\mu\left(\frac{G}{F_S}\right)\sigma^\mu\left(\frac{\bar{G}}{\bar{F}_S}\right)(\partial A)^2+h.c.\right]\\
     &+\left[fF_S-\frac{i}{2}f'\frac{G\sigma^\mu\bar{G}}{\bar{F}_S}\partial_\mu A-\frac{g''}{2}\left(\frac{\bar{G}}{\bar{F}_S}\right)^2(\partial A)^2-\frac{g'}{2}\left(\frac{\bar{G}}{\bar{F}_S}\right)^2\Box A\right.\\
    &\left.\qquad-g'\partial_\nu\left(\frac{\bar{G}}{\bar{F}_S}\right)\bar{\sigma}^\mu\sigma^\nu\frac{\bar{G}}{\bar{F}_S}\partial_\mu A +h.c.\right].
    \end{aligned}
\label{eqn:o_LKW}
\end{equation}
The next step is to integrate out the auxiliary field $F_S$, which means that we need to find the (algebraic) solution to its EoM and then plug it back into lagrangian (\ref{eqn:o_LKW}). Such solution can be found iteratively as an expansion in the number of fermions. Among the operators which couple the goldstino to the scalar, we only keep those with at most two goldstini, i.e. we apply the quadratic limit.  Putting everything together and performing integrations by parts, the resulting lagrangian is
\begin{equation}
    \begin{aligned}
    \mathcal{L}=&-|f|^2+\frac{h}{2}\partial_\mu A\partial^\mu A+\left(\frac{i}{2}\partial_\mu G \sigma^\mu\bar{G}+h.c.\right)+\(i\frac{f'}{2f}+h.c.\)G\sigma^\mu\bar{G}\partial_\mu A\\
    &+\frac{h}{2|f|^2}\left[i\,\partial^\mu G \sigma^\nu\bar{G}\partial_\mu A\partial_\nu A-\frac{i}{2}\partial_\mu G \sigma^\mu\bar{G}\partial_\nu A\partial^\nu A+h.c.\right]-\frac{1}{4|f|^2}\bar{G}^2\Box G^2\\
    &+\left[\frac{\bar{g}'}{\bar{f}^2}\left(\frac{1}{2} G^2\Box A+\partial_\mu G \sigma^\mu\bar{\sigma}^\nu G \partial_\nu A\right)+\xi\, G^2\,\partial_\mu A\partial^\mu A+h.c.\right] \ ,
    \end{aligned}
\label{eqn:o_L_tot}
\end{equation}
where
\begin{equation}
    \xi\equiv \frac{\bar{g}''}{2\bar{f}^2}-\frac{\bar{g}'\bar{f}'}{\bar{f}^3} \ .
\label{eqn:o_xi_def}
\end{equation}
From this lagrangian, one can extract the goldstino sound speed, and compute $2\to 2$ scattering amplitudes in order to extract positivity bounds from them. We present such results below, but, as announced in the beginning, we first perform some field redefinitions, in order to remove the redundant 4-fields operators and also to exchange most of the 3-fields operators with 4-fields ones: this makes manifest the proper combination of coefficients subject to positivity bounds.  In (\ref{eqn:o_L_tot}) there are three operators which are suitable to this procedure (i.e. which are proportional to the free EoM),
\beq
\partial_\mu G \sigma^\mu\bar{G}\partial_\nu A\partial^\nu A\ , \quad \partial_\mu G \sigma^\mu\bar{\sigma}^\nu G \partial_\nu A\ , \quad G^2\Box A
\eeq
(plus hermitian conjugates), which can be removed with the following field redefinitions, respectively,
\begin{equation}
        \delta G^\alpha =\frac{h}{4|f|^2}(\partial A)^2\,G^\alpha \ , \quad \delta G^\alpha=-i\frac{g'}{f^2}(\bar{G}\bar{\sigma}^\mu)^\alpha\,\partial_\mu A \ , \quad \delta A=\frac{\bar{g}'}{2h\bar{f}^2}G^2+h.c \ .
\eeq
Eventually, the final resulting lagrangian in the four-field limit is
\begin{equation}
    \begin{aligned}
        \mathcal{L}=&-|f|^2+\frac{h}{2}\partial_\mu A\partial^\mu A+\(\frac{i}{2}\partial_\mu G \sigma^\mu\bar{G}+h.c.\)+\(mG^2+h.c.\)+\lambda G^2\bar{G}^2\\
        &+\(i\frac{f'}{2f}G\sigma^\mu\bar{G}\partial_\mu A+\tilde{\xi}G^2\partial_\mu A\partial^\mu A+h.c.\)+\frac{1}{8h}\left(\frac{\bar g'^2}{\bar f^4}G^2\Box G^2+h.c.\right)\\
        &+\frac{1}{2|f|^2}\left(h-\frac{2|g'|^2}{|f|^2}\right)\left(i\,\partial^\mu G \sigma^\nu\bar{G}\partial_\mu A\partial_\nu A+h.c.\right)-\frac{1}{4|f|^2}\left(1-\frac{|g'|^2}{h|f|^2}\right)\bar{G}^2 \Box G^2 \ ,
    \end{aligned}
\label{eqn:o_L_final}
\end{equation}
where $m$, $\lambda$ and $\tilde{\xi}$ denote combinations of original coefficients which are not needed explicitly. Note that from \eqref{eqn:o_L_final}, one cannot compute the goldstino sound speed exactly, but only up to terms involving two derivative in the scalar, as a result of the four-field limit.

\subsubsection{Positivity bounds and sound speed}

 In this paper, we focus on (twice-subtracted) $2\to 2$ amplitudes in the forward limit ($p_1=p_3$), which are subject to direct positivity bounds. Although they do not capture the full set of all positivity bounds in a generic lagrangian, they suffice to identify an amplitude version of the breakdown of causality discussed in Section~\ref{section:equivalence}. In the lagrangian \eqref{eqn:o_L_final}, they concern the following amplitudes,
   \beq
 \bead
\cA(G,\bar G\to G,\bar G)&=\frac{h|f|^2-|g'|^2}{h|f|^4}[14]\langle 23 \rangle\(p_1-p_4\)^2=\frac{h|f|^2-|g'|^2}{h|f|^4}s^2 \\
\cA(G, G\to \bar G, \bar G)&=\frac{\bar g'^2}{h\bar f^4}\([12][34](s-t)+[14][23](u-t)\)=\frac{2\bar g'^2}{h\bar f^4}
s^2 \ , \\
\cA(G,A\to G,A)&= -\frac{h|f|^2-2|g'|^2}{2|f|^4}\(p_1+p_3\)^\mu [1\(p_{2,\mu}\slashed{p}_4+p_{4,\mu}\slashed{p}_2\)3\rangle=\frac{h|f|^2-2|g'|^2}{2|f|^4}s^2 \ ,
\eead
\eeq
 using notations of the massless helicity spinor formalism (see e.g. \cite{Elvang:2013cua}), and where the last equalities hold in the forward limit. We dropped all terms which scale differently than $s^2$ in the forward limit, since they do not contribute to the twice-subtracted bounds. By construction, the factors match the coefficients appearing in the lagrangian (\ref{eqn:o_L_final}). Of course, they could have been obtained directly from \eqref{eqn:o_L_tot}, since amplitudes do not depend on the operator basis used. Other amplitudes, corresponding to crossed or time-reversed versions of those above, do not bring further constraints. 
 
 Following \cite{Bellazzini:2016xrt, Dine:2009sw}, we can derive a positivity bound on the elastic amplitude $\cA(G,\bar G\to G,\bar G)$, which reads (restoring now the $A$-dependence of the various functions)
\begin{equation}
    \frac{d^2 \cA(G,\bar G\to G,\bar G)}{ds^2}\Bigg\vert_{t=s=0}\geq 0 \iff h(A)|f(A)|^2>|g'(A)|^2 \ ,
\end{equation}
i.e. the coefficient of the operator $\bar{G}^2\Box G^2$ must be negative when there are no three-point couplings contributing to the twice-subtracted amplitudes. An even stronger bound follows from including the inelastic amplitude $\cA(G, G\to \bar G,\bar G)$ \cite{Bellazzini:2014waa} (see in particular \cite{Trott:2020ebl} for the Weyl fermion case), 
\begin{equation}
\label{eq:inelasticGoldstinoScattering}
    \frac{d^2 \cA(G,\bar G\to G,\bar G)}{ds^2}-\frac{1}{2}\abs{\frac{d^2 \cA(G,G\to \bar G,\bar G)}{ds^2}}\,\Bigg\vert_{t=s=0}\geq 0 \iff h(A)|f(A)|^2>2|g'(A)|^2 \ .
\end{equation}
This bound combines the coefficients of the $\bar{G}^2\Box G^2$ and $G^2\Box G^2$ operators, when no three-point coupling contributes to the twice-subtracted amplitudes. From $\cA(G,A\to G,A)$, one gets
\begin{equation}
	h(A)|f(A)|^2>2|g'(A)|^2 \ ,
\label{eqn:o_pos_bound}
\end{equation}
i.e. the coefficient of $\left(i\,\partial^\mu G \sigma^\nu\bar{G}\partial_\mu A\partial_\nu A+h.c.\right)$ must be positive when no three-point coupling contributes to the twice-subtracted amplitudes. (\ref{eqn:o_pos_bound}) immediately yields the strongest bound among the two derived from pure Goldstino scattering, and including the inelastic amplitude $\cA( G,A\to \bar G,A)$ does not improve it, as it vanishes in the theory under consideration when twice-subtracted, or in the forward limit\footnote{There also exists constraints beyond forward scattering or two subtractions \cite{Caron-Huot:2020cmc}, whose applications to our models are left for future study.}.

Strictly speaking, the positivity bounds only apply to $2\to 2$ amplitudes in the vacuum, which we take to be $A=0$ without loss of generality. Therefore, they constrain only the first order coefficients of the dangerous operators, i.e.
\begin{equation}
    h(0)|f(0)|^2>2|g'(0)|^2 \ .
\label{eqn:pos_bound_zero}
\end{equation}
The higher orders of the Taylor expansion of the scalar functions $f,g',h$ carry higher powers of the scalar field $A$ and thus contribute to amplitudes of higher multiplicity. There may exist constraints on higher-multiplicity amplitudes which can effectively be resumed to generate (\ref{eqn:o_pos_bound}), but we are not aware that they have already been precisely identified and derived. Therefore, we remain conservative and only claim that (\ref{eqn:pos_bound_zero}) follows from scattering-amplitudes positivity. Nevertheless, we argue here that we need to extend the bound (\ref{eqn:pos_bound_zero}) to the whole functions' domain in order to allow for a time-dependent dynamics of the field $A$. Considering the case of a time-dependent solution $A=A(t)$ (relevant or not for an inflationary scenario), this fact can be seen by studying the sound speed of the goldstino, as discussed in Section \ref{section:equivalence}, which now becomes non-trivial because of the presence of the higher-order operators generated by the constraints. The sound speed can be computed from the quadratic goldstino EoM on the $A(t)$ background. The most suitable form of the lagrangian to carry out this computation is actually (\ref{eqn:o_L_tot}), since it captures exactly all the operators involving two goldstini, which are the only ones contributing to the computation of the sound speed. However, unlike in (\ref{eqn:o_L_final}), quadratic terms in the third line of \eqref{eqn:o_L_tot} contribute to the quadratic goldstino EoM and need to be dealt with\footnote{\label{footnotef'}We do not need to consider the term proportional to $f'$. Indeed, it vanishes for real $f$, which can be achieved by redefining the goldstino and its auxiliary field, i.e. by decomposing $f(A)=\abs{f(A)}e^{i\alpha(A)}$ and performing $(G,F_S)\to e^{-i\alpha(A)}(G,F_S)$ in \eqref{eqn:o_LKW}. In \eqref{eqn:o_LKW} this changes $f\to \abs{f}$ and the term in the third line proportional to $\frac{f'}{\bar F_S}$ into $i\alpha'\(\frac{\abs{f}}{\bar F_S}+1\)$, which vanishes on the bosonic EoM of $F_S$. Therefore, one can consider $f\in\mathbb{R}$ when computing the goldstino sound speed.}. Following \cite{Kolb:2021xfn}, we write the quadratic goldstino EoM, 
\beq
i\(1+\frac{h\dot A{}^2}{2\abs{f}^2}\) \partial_0 G \sigma^0
+ i\(1-\frac{h\dot A{}^2}{2\abs{f}^2}\)\partial_i G \sigma^i + 2  \dot A\frac{\bar g'}{\bar f^2} \partial_i \bar G \bar \sigma^i = 0 \ ,
\eeq
which we Fourier transform, using rotation invariance to write the momentum as $k^\mu=(k_0,0,0,k_3)$,
\beq
\(1+\frac{h\dot A{}^2}{2\abs{f}^2}\)k_0G(k) \sigma^0 
+ \(1-\frac{h\dot A{}^2}{2\abs{f}^2}\)k_3 G(k) \sigma^3 + 2 i \dot A\frac{\bar g'}{\bar f^2}k_3 \bar G(k) \bar \sigma^3 = 0 \ .
\eeq
We then decompose along the spinor components of $G^\alpha=\(G^+ \ G^-\)$,
\beq
\(1+\frac{h\dot A{}^2}{2\abs{f}^2}\)k_0G^\pm  
-(\pm) \(1-\frac{h\dot A{}^2}{2\abs{f}^2}\)k_3 G^\pm + 2 i \dot A\frac{\bar g'}{\bar f^2}k_3 \bar G^\mp = 0 \ ,
\eeq
and plug the EoM of $G^\pm$ into that of $G^\mp$ to get
\beq
\left\{\(1+\frac{h\dot A{}^2}{2\abs{f}^2}\)^2k_0^2  -\[\(1-\frac{h\dot A{}^2}{2\abs{f}^2}\)^2+4 \dot A{}^2\frac{\abs{g'}^2}{\abs{f}^4}\]k_3^3\right\} G^\pm = 0 \ .
\eeq
From this, one immediately reads
\beq
c_s^2= 1 - \frac{4\dot A{}^2}{\(\abs{f(A)}^2+\frac{h(A)\dot A{}^2}{2}\)^2} \(\frac{h(A)}{2}\abs{f(A)}^2 -\abs{g'(A)}^2\) \ .
\label{eqn:o_cs}
\eeq
When $h=1$, $g'=0$, one recovers the result of \cite{Kahn:2015mla}. This formula shows that, in order to always have a causal/subluminal propagation of the gravitino, we need to extend the positivity bound (\ref{eqn:pos_bound_zero}) to the whole field space, i.e. (\ref{eqn:o_pos_bound}). 

This result is coherent with the discussion in \cite{Dine:2009sw} and it agrees with the study of longitudinal gravitino dynamics in SUGRA \cite{Kolb:2021xfn, Dudas:2021njv}. As already anticipated, we explicitly determined that the low-energy goldstino sound speed (\ref{eqn:o_cs}) exactly matches the gravitino's one in SUGRA (\ref{eq:orthog4}),
since the latter one has no Planck-suppressed corrections. We are not aware of a deep reason for the absence of supergravity corrections in this case. Being the two sound speeds identical, the causality requirement (\ref{eqn:o_pos_bound}) in the low-energy theory directly matches the subluminality one in the SUGRA framework.

The causality constraint  (\ref{eqn:o_pos_bound})  implies that the functions defining the effective theory in (\ref{eq:orthog2}) are not arbitrary. They need to satisfy an additional condition for all points in the theory parameter space reached during the time-dependent dynamics, in order to define a consistent theory. This is a signal that the starting point, namely the orthogonal constraint (\ref{eqn:o_const}), may not be well-defined microscopically. Indeed, the UV origin of such constraint is not clear; it seems to be determined by a lagrangian which already contains higher-derivative operators \cite{DallAgata:2016syy}. We will present in later sections a different viewpoint on this issue and show that specific microscopic higher-derivative operators remove the causality constraint and render therefore the model causal in all points of the theory parameter space.

\subsection{Light complex scalar}
\label{section:complex}

The second theory we study is very similar to the previous one, with the difference that we have a complex scalar instead of a real one. Again, we consider the goldstino superfield $S$ (\ref{eqn:goldstino_sf}) coupled to another chiral superfield,
\begin{equation}
	\mathcal{H}= H + \sqrt{2}\theta \chi_H +\theta^2 F_H,
\end{equation} 
but this time the constraint is 
\begin{equation}
	S\bar{\mathcal{H}}=\text{chiral},
\label{eqn:cs_const}
\end{equation}
which fixes \cite{Komargodski:2009rz}
\beq
	\chi_H=i\sigma^\mu\frac{\bar{G}}{\bar{F}_S}\partial_\mu H \ ,\quad 	F_H=-\partial_\nu\left(\frac{\bar{G}}{\bar{F}_S}\right)\bar{\sigma}^\mu\sigma^\nu \frac{\bar{G}}{\bar{F}_S} \partial_\mu H-\frac{1}{2}\left(\frac{\bar{G}}{\bar{F}_S}\right)^2\Box H \ . 
	\label{eqn:complexSolution}
\eeq
The most general model is characterised by the K\"ahler potential and superpotential
of \eqref{eqn:cs_superpot}. The off-shell lagrangian  turns out to be
\begin{equation}
	\begin{aligned}
		\mathcal{L}=&\, \xi F_S\bar{F}_S +\xi\left(\frac{i}{2}\partial_\mu G \sigma^\mu {\bar G}+h.c.\right)+\kappa_{H \bar H}\partial_\mu H \partial^\mu {\bar H}\\
		&+\frac{\kappa_{H\bar H}}{2}\left[i\partial^\mu \(\frac{G}{F_S}\)\sigma^\nu \(\frac{\bar{G}}{\bar{F}_S}\)+h.c.\right]\left(\partial_\mu {\bar H} \partial_\nu H+\partial_\mu H\partial_\nu {\bar H} \right)\\
		&-\frac{\kappa_{H\bar{H}}}{2}\[i\partial_\mu\( \frac{G}{F_S}\)\sigma^\mu\(\frac{\bar{G}}{\bar{F}_S}\)+h.c.\]\partial_\nu H\partial^\nu\bar{H}\\
	&+\frac{\kappa_{H \bar H}}{2}\frac{G \sigma^\mu \bar G}{\abs{F_S}^2}\left(i\partial_\mu H\Box {\bar H} +h.c.\right)-\frac{\xi}{4}\left(\frac{\bar{G}^2}{\bar{F}_S}\right)\Box\left(\frac{G^2}{F_S}\right)\\
	&-\kappa_{H \bar H} \epsilon^{\mu\nu\alpha\beta} \partial_\mu\(\frac{G}{F_S}\)\sigma_\lambda \(\frac{\bar{G}}{\bar{F}_S}\)\partial_\alpha H \partial_\beta {\bar H} \\
	&+\left\{f(H) F_S - \left(\frac{\xi_H}{2}+\frac{f'}{\bar{F}_S}\right)i G \sigma^\mu {\bar G} \partial_\mu H +\frac{g''(H)}{2} \(\frac{\bar{G}}{\bar{F}_S}\)^2 \left(\partial H\right)^2\right.\\
	&\left.\,\,\,\,\,+\frac{g'(H)}{2}\(\frac{\bar{G}}{\bar{F}_S}\)^2 \Box H+g'(H) \(\frac{\bar{G}}{\bar{F}_S}\) \bar{\sigma}^\mu\sigma^\nu\partial_\nu\(\frac{\bar{G}}{\bar{F}_S}\)\partial_\mu H+h.c.\right\} \ .
	\end{aligned}
\label{eqn:cs_L_off}
\end{equation}
It was argued in \cite{DallAgata:2015zxp} that one can set $\xi=1$ from the very beginning without loss of generality, which can be seen explicitly from the lagrangian above: the whole $\xi$-dependence can be redefined away, up to terms with four goldstini and at least two scalar fields which are irrelevant in this paper, by the rescaling
\beq
(G,F_S) \longrightarrow \frac{1}{\sqrt{\xi}}(G,F_S) \ , 
\eeq
which also implies $s\to \xi^{-1/2}s$ for the constrained sgoldstino. Actually, $S\to \xi^{-1/2}S$ still defines a chiral nilpotent superfield $S$, due to the constraints on $S$ and $H$. After redefining $f(H)\to\sqrt{\xi(H,\bar H)}f(H,\bar H)$ (where $f$ is no longer holomorphic), $\xi=1$ disappears from the action\footnote{After all the redefinitions but before integrating $F_S$, one finds a leftover $\xi$-dependent term,
\bes
-\frac{\xi_H}{2\xi}\(\frac{f}{\bar F_S}+1\)iG\sigma^\mu\bar G\partial_\mu H+h.c. \ ,
\ees
which however becomes quartic in goldstini with more than four fields once $F_S$ is put on-shell.}.

After integrating out the auxiliary field, the resulting on-shell lagrangian becomes
\begin{equation}
\begin{aligned}
	\mathcal{L}=&-\abs{f
	}^2+\left(\frac{i}{2}\partial_\mu G \sigma^\mu {\bar G}+h.c.\right)+\kappa_{H\bar H}\partial_\mu H\partial^\mu {\bar H}\\
	&-\frac{1}{4\abs{f}^2} {\bar G}^2 \Box G^2+\frac{\kappa_{H \bar H}}{2\abs{f}^2}\left(i\partial^\mu G \sigma^\nu {\bar G}+h.c.\right)\left(\partial_\mu H\partial_\nu {\bar H} +\partial_\nu H \partial_\mu {\bar H} \right)\\
	&-\frac{\kappa_{H\bar{H}}}{2|f|^2}\(i\partial_\mu G\sigma^\mu\bar{G}+h.c.\)\partial_\nu H\partial^\nu\bar{H}+\left[i\frac{f
	_H}{f} G \sigma^\mu {\bar G} \partial_\mu H+h.c.\right]\\
	&+\frac{\kappa_{H \bar H}}{\abs{f}^2}\[\frac{1}{2}G \sigma^\mu {\bar G} \left(i\,\partial_\mu H\Box {\bar H} +h.c.\right)-\epsilon^{\mu\nu\alpha\beta}\partial_\mu G \sigma_\nu {\bar G}  \partial_\alpha H\partial_\beta {\bar H}\]\\
	&+\left[\frac{g''}{2f^2} {\bar G}^2(\partial H)^2+\frac{g'}{2f^2} {\bar G}^2\Box H+\frac{g'}{f^2} {\bar G} \bar{\sigma}^\mu\sigma^\nu\partial_\nu {\bar G}  \partial_\mu H +h.c.\right] \ .
\end{aligned}
\label{eqn:cs_L_on}
\end{equation}
As in the previous section, this is the lagrangian corresponding to what we called the quadratic limit. We can move to the four-fields limit by performing the following field redefinitions:
\begin{equation}
\begin{aligned}
    \delta G^\alpha=&\frac{\kappa_{H\bar{H}}}{2 |f|^2}\partial_\mu H\partial^\mu\bar{H}G^\alpha,  &&&&& \delta H=&\frac{i}{2|f|^2}G\sigma^\mu\bar{G}\partial_\mu H, \\
    \delta G^\alpha=& -i\frac{g'}{f^2}({\bar G} \bar{\sigma}^\mu)^\alpha\,\partial_\mu H, &&&&& \delta H=&\frac{\bar{g}'}{2\kappa_{H \bar H}\bar{f}^2} G^2.
\end{aligned}
\end{equation}
The transformations in the first line remove the redundant operators $\(i\partial_\mu G\sigma^\mu\bar{G}+h.c.\)\partial_\nu H\partial^\nu\bar{H}$  and $\(G\sigma^\mu\bar{G}\partial_\mu H+h.c.\)$ from the lagrangian, while those in the second one transform respectively the 3-fields operators $\(\bar G^2 \Box H+h.c.\)$
and $\(\bar G \bar{\sigma}^\mu\sigma^\nu\partial_\nu {\bar G} \partial_\mu H +h.c.\)$ into 4-fields ones. The resulting lagrangian is
\begin{equation}
\begin{aligned}
	\mathcal{L}=&-\abs{f}^2+\kappa_{H \bar H}\partial_\mu H\partial^\mu {\bar H} +\left(\frac{i}{2}\partial_\mu G \sigma^\mu {\bar G} +h.c.\right)-\left(m_1\, G^2+h.c.\right)\\
	&+2\left[\bar{m}_1({\bar k} {\bar H}-kH) {\bar G}^2+h.c.\right]-\frac{1}{4\abs{f}^2}\left(1-\frac{\abs{g'}^2}{\abs{f}^2\kappa_{H\bar H}}\right) {\bar G}^2\Box G^2\\
	&+\frac{1}{2}\left(k^2H^2+{\bar k}^2 {\bar H}^2-2\abs{k}^2\abs{H}^2\right)\left(m_1 G^2+\bar{m}_1 {\bar G}^2\right)-m_2\,G^2 {\bar G}^2
	\\
	&+\frac{1}{2\abs{f}^2}\left(\kappa_{H \bar H}+\frac{\abs{g'}^2}{\abs{f}^2}\right)\(i\partial^\mu G \sigma^\nu\bar{G}\partial_\mu H\partial_\nu {\bar H}+h.c.\)\\
	&+\frac{1}{2\abs{f}^2}\left(\kappa_{H \bar H}-3\frac{\abs{g'}^2}{\abs{f}^2}\right)\(i\partial^\mu G\sigma^\nu\bar{G}\partial_\mu {\bar H} \partial_\nu H+h.c.\)\\
	&+\left[\lambda_1 G^2\partial_\mu H\partial^\mu {\bar H}+\lambda_2 {\bar G}^2(\partial H)^2+h.c.\right]+\(\frac{i\(2f_H\bar f-\partial_H\abs{f}^2\)}{2\abs{f}^2}G\sigma^\mu {\bar G}\partial_\mu H+h.c. \) \\
	&-\frac{1}{\abs{f}^2}\left(\kappa_{H \bar H}+\frac{\abs{g'}^2}{\abs{f}^2}\right)\epsilon^{\mu\nu\alpha\beta}\partial_\mu G \sigma_\nu {\bar G} \partial_\alpha H\partial_\beta {\bar H},
\end{aligned}
\label{eqn:cs_L_tot}
\end{equation}
where $m_1$, $m_2$, $\lambda_1$ and $\lambda_2$ are a short-hand notation for coefficients depending on $H$ whose explicit form is not needed in what follows.

\subsubsection{Positivity bounds and sound speed}

We can now present the positivity bounds on the coefficients of lagrangian (\ref{eqn:cs_L_tot}) (although we stress again that the bounds do not depend on a precise choice of operator basis). As already said, they are obtained from (twice-subtracted) $2\to 2$ scattering amplitudes in the forward limit, to which only the operators
$ \partial^\mu G \sigma^\nu {\bar G} \partial_\mu H\partial_\nu {\bar H}$, its $H\leftrightarrow \bar H$ version and ${\bar G}^2\Box G^2$ contribute. Unlike the previous section, the operator $G^2\Box G^2$ is absent, and decomposing $H=A+iB$, the two-fermion two-scalar operators can be further written as
\beq
\bead
\frac{1}{2\abs{f}^2}&\left\{i\partial^\mu G \sigma^\nu {\bar G} \left[\left(\kappa_{H \bar H}+\frac{\abs{g'}^2}{\abs{f}^2}\right)\partial_\mu H\partial_\nu {\bar H} +\left(\kappa_{H \bar H}-3\frac{\abs{g'}^2}{\abs{f}^2}\right)\partial_\mu {\bar H} \partial_\nu H\right]+h.c.\right\}=\\	
&=\frac{1}{\abs{f}^2}\left(\kappa_{H \bar H}-\frac{\abs{g'}^2}{\abs{f}^2}\right)\left(\partial^\mu G \sigma^\nu {\bar G} +h.c.\right)(\partial_\mu A\partial_\nu A+\partial_\mu B\partial_\nu B)\\
&\quad+2\frac{\abs{g'}^2}{\abs{f}^4}\left[\partial^\mu G \sigma^\nu {\bar G}  (\partial_\mu A\partial_\nu B-\partial_\nu A\partial_\mu B)+h.c.\right] \ , \label{eq:causality-complexsc}
\eead
\eeq
where the last line does not generate any non-trivial inelastic amplitude $\cA(G, A \to G, B)$ in the forward limit.
The positivity bounds can therefore be extracted from those on elastic scattering in the real scalar case. In this case, this suffices for all operators to yield the same bound,
\begin{equation}
	\kappa_{H \bar H}(0)\abs{f(0)}^2>\abs{g'(0)}^2 \ .
\label{eqn:cs_pos_bound_0}
\end{equation}
Following the same reasoning of the real scalar case, we motivate the need for the bound (\ref{eqn:cs_pos_bound_0}) to hold in the whole field space, and not just in the vacuum, by studying the goldstino sound speed. Using the quadratic limit of the lagrangian (\ref{eqn:cs_L_on}) and considering a time-dependent scalar background $H=H(t)$, the goldstino sound speed can be computed, following the same steps\footnote{Also in this case the operator $\(G\sigma^\mu\bar{G}\partial_\mu H+h.c.\)$ does not actually contribute to the sound speed. The argument is the same as in footnote \ref{footnotef'}, adapted to the complex scalar case. More precisely, in \eqref{eqn:cs_L_tot}, in the combination $\(2f_H\bar f-\partial_H\abs{f}^2\)\partial_\mu H$, the only non-vanishing terms are $\(2f_H\bar f-\partial_H\abs{f}^2\)\partial_\mu H/\xi^2$, where now $f$ is again the holomorphic function of \eqref{eqn:cs_L_off}. This whole expression can be rewritten as $ \bar f\partial_\mu f/\xi^2$, and one can absorb the phase of $f$ in $G$, as in footnote \ref{footnotef'}.} that lead to (\ref{eqn:o_cs}), to be
\begin{equation}
    c_s^2=1-\frac{4|\dot H|^2}{\(|f|^2+\kappa_{H\bar{H}}|\dot H|^2\)^2}\left[\kappa_{H\bar{H}}|f|^2-|g'|^2\right] \ ,
\label{eqn:cs_cs}
\end{equation}
so that subluminality enforces the causality bound (\ref{eqn:cs_pos_bound_0}) to be respected for any time-dependent solution and therefore for any value of the scalar field scanned by such solutions,
\begin{equation}
    \kappa_{H \bar H}\abs{f}^2>\abs{g'}^2 \ .
\label{eqn:cs_pos_bound}
\end{equation}
Notice that, as expected, the (improved) condition (\ref{eqn:cs_pos_bound}), as well as the sound speed (\ref{eqn:cs_cs}), agree with the one coming from the subluminality of the gravitino sound speed in SUGRA  (\ref{eqn:cs_sublu}) in the decoupling limit $M_P \to \infty$. As discussed in Section \ref{section:equivalence}, the plausible interpretation of the gravitational corrections in the SUGRA condition is that the positivity of the four-field operators can be violated by Planck suppressed $1/M_P^2$ terms.
Causality should nevertheless hold or be unresolvable \cite{Alberte:2020jsk}. The occurence of positivity bounds suggests, like for the orthogonal constraint,  that the superfield constraint (\ref{eqn:cs_const}) does not have a two-derivative microscopic origin. The problem seems again that the auxiliary field is determined by the constraint, and this has no obvious microscopic  interpretation in terms of decoupling of physical degrees of freedom in a two-derivative lagrangian.

\subsection{Light fermion}
\label{section:fermion}

For completeness, we consider briefly a theory in which the component to be removed is the complex scalar one. This situation corresponds to giving a large mass to a complex scalar in a chiral multiplet and is straightforward to realize microscopically, up to fine-tuning. Consequently we expect no non-trivial causality condition on the low-energy goldstino lagrangian.   Calling the chiral superfield $\mathcal{Q}=Q+\sqrt{2}\theta\chi_Q+\theta^2F_Q$, the decoupling of the scalar $Q$ is achieved via the constraint
\begin{equation}
	S\mathcal{Q}=0 \ ,
\label{eqn:df_const}
\end{equation}
which, similarly to the orthogonal constraint case (\ref{eqn:o_const}), implies $\mathcal{Q}^3=0$ and fixes\footnote{For more specific discussions on other types of constraints one can see for example \cite{Dudas:2011kt, Kuzenko:2017oni, Aldabergenov:2021obf}.} \cite{Komargodski:2009rz}
\begin{equation}
	Q=\frac{\chi_Q G}{F_S}-\frac{F_Q}{2F_S^2}\, G^2 \ .
\label{eqn:df_Q_expr}
\end{equation}
It can be shown that the most general theory is given by a superpotential which is analogous to the previous ones,
\begin{equation}
	W=f(Q)S+g(Q) \ ,
\label{eqn:dF_Superpot}
\end{equation}
and by the K\"ahler potential
\begin{equation}
	K=S\bar{S}+Q\bar{Q}+\frac{1}{2}Q\bar{Q}\left(h_1Q+\bar{h}_1\bar{Q}\right)+\frac{h_2}{4}\left(Q\bar{Q}\right)^2 \ ,
\label{eqn:df_K_pot}
\end{equation}
with $h_1$ and $h_2$ constants, respectively complex and real.

By writing explicitly the on-shell lagrangian, we find that the coefficients of the operators subject to positivity bounds have the appropriate signs for any values of the theory space parameters, implying that there are no causality constraints to be imposed. Therefore, unlike the two previous cases, the constraint (\ref{eqn:df_const}) removing the scalar component of the superfield does not demand that we impose any extra condition in order for the resulting theory to be causal. Since (\ref{eqn:df_const}) leaves the auxiliary field $F_Q$ unconstrained, this is  further evidence that the causality problems arise only when an auxiliary field is removed, as pointed out at the end of the last section. One could object that the problems could come from the removal of a fermion, which is also implied by the constrains of Sections~\ref{section:orthogonal} and \ref{section:complex}, but it turns out not to be the case, as we will show in the next sections. Indeed, decoupling a fermion can be achieved microscopically from a two-derivative lagrangian, albeit modulo a fine-tuning. The tricky constraints are those eliminating auxiliary fields, which cannot be done, in our opinion, without using higher-derivative operators directly in the UV, as already anticipated in \cite{DallAgata:2016syy}. 

\section{Evading causality conditions with alternative constraints }\label{section:twoconstraints}

In the previous sections, we have studied models realized by various superfields constraints and showed that  the ones that affect the auxiliary field component of a chiral superfield lead to nontrivial positivity bounds, which restricts the parameter space of the theory and may indicate that the corresponding UV theory is not a two-derivative one.

Now we go deeper into this analysis, making use of the generalized superfield constraint formalism developed in \cite{DallAgata:2016syy}, where it was shown that one can remove the lowest component of a superfield $Q_\textup{L}$ (where $L$ denotes a possible Lorentz index) by applying the constraint
\begin{equation}
	S\bar{S}Q_\textup{L}=0 \ .
\label{eqn:gen_const_def}
\end{equation} 
This is solved, implicitly and in superspace, by
\begin{equation}
	Q_\textup{L}=-2\frac{\bar{D}_{\dot{\alpha}}\bar{S}\bar{D}^{\dot{\alpha}}Q_\textup{L}}{\bar{D}^2\bar{S}}-\frac{\bar{S}\bar{D}^2Q_\textup{L}}{\bar{D}^2\bar{S}}-2\frac{D^\alpha SD_\alpha\bar{D}^2\left(\bar{S}Q_\textup{L}\right)}{D^2S\bar{D}^2\bar{S}}-S\frac{D^2\bar{D}^2\left(\bar{S}Q_\textup{L}\right)}{D^2S\bar{D}^2\bar{S}} \ .
\label{eqn:gen_const_sol}
\end{equation}
The only nontrivial constraint is actually on the lowest component (i.e. $\theta=\bar{\theta}=0$) of $Q_L$. The power of the constraint (\ref{eqn:gen_const_def}) is that it allows to remove one single component at a time, and this means that every superfields constraint can be expressed as a combination of multiple such single, generalized constraints. For instance, the orthogonal constraint (\ref{eqn:o_const}) can be decomposed into the three following single constraints \cite{DallAgata:2016syy},
\begin{align}
	S\bar{S}\left(\Phi-\bar{\Phi}\right)=&0, \label{eqn:gc_sc}\\
	S\bar{S}D_\alpha\Phi=&0,\label{eqn:gc_ferm}\\
	S\bar{S}D^2\Phi=&0, \label{eqn:gc_aux}
\end{align}
of which the first one removes the imaginary part of the scalar field component of $\Phi$, the second one the fermionic one and the third one the auxiliary field one, reproducing equation (\ref{eqn:orthogonalSolution}).

\subsection{Removing the imaginary scalar and the fermion}

Focusing on this set of constraints, we see that by using simultaneously (\ref{eqn:gc_sc}) and (\ref{eqn:gc_ferm}) one obtains a theory with the same physical degrees of freedom as the orthogonal constraint case. The crucial difference is that here we will not impose (\ref{eqn:gc_aux}), so that the auxiliary field $F_{\phi}$ will be determined by its usual algebraic EoM from the off-shell lagrangian: with the issues regarding the removal of the inflaton auxiliary field being absent (now $F_\phi$ is unconstrained), we believe that the resulting theory should represent a minimal inflationary model in SUGRA, since, according to the discussion up to this point, it is likely to have a well-understood origin in terms of a SUSY-breaking UV theory.

The solution of the two single, generalized constraints (\ref{eqn:gc_sc}) and (\ref{eqn:gc_ferm}) is displayed in Appendix \ref{appendix:solutionsConstraints}. Instead, we present here the solution of the two constraints combined together\footnote{This is obtained by solving - at the appropriate order in the number of fields - the system given by the two implicit equations (\ref{eqn:gc_eq_B}) and (\ref{eqn:gc_eq_psi}) of Appendix \ref{appendix:solutionsConstraints}.}, restricted ourselves to the appropriate number of the fields that will lead to at most four fields in the final action:
\begin{align}
    B=&\frac{i}{4}\[\(\frac{\bar{G}}{\bar{F}_S}\)^2\bar{F}_\phi-\(\frac{G}{F_S}\)^2F_\phi\]+\frac{G\sigma^\mu\bar{G}}{2|F_S|^2}\partial_\mu A +\dots \ ,\\
    \psi_\alpha=&\frac{F_\phi}{F_S}G_\alpha+i\frac{\(\sigma^\mu\bar{G}\)_\alpha}{\bar{F}_S}\partial_\mu A-\frac{i}{4}\frac{\(\sigma^\mu\bar{G}\)_\alpha}{\bar{F}_S}\partial_\mu\[\(\frac{G}{F_S}\)^2F_\phi\]+\dots \ ,
\end{align}
where "$\dots$" denotes terms irrelevant for the quadratic of four-field limits in the final lagrangian, which are not important for our considerations. 

The most general theory associated to this new setup is characterized by
\beq
    K=S\bar{S}+\kappa(\Phi,\bar{\Phi})\ , \quad W=f(\Phi)S+g(\Phi) \ ,
\eeq
where $\kappa(\Phi,\bar{\Phi})$ is a real function and $f(\Phi)$ and $g(\Phi)$ are instead holomorphic. The resulting full off-shell lagrangian is
\begin{equation}
    \begin{aligned}
        \mathcal{L}=&|F_S|^2+\kappa_{\phi\bar{\phi}}|F_\phi|^2+\kappa_{\phi\bar{\phi}}\partial_\mu A\partial^\mu A+\(\frac{i}{2}\partial_\mu G\sigma^\mu \bar{G}+h.c.\)+\[fF_S+g'F_\phi+h.c.\]\\
        &-\frac{\kappa_{\phi\bar{\phi}\phi}+\kappa_{\phi\bar{\phi}\bar{\phi}}}{4}\[\(\frac{G}{F_S}\)^2F_\phi^2\bar{F}_\phi+h.c.\]-\frac{1}{4}\(f'F_S+g''F_\phi+h.c.\)\[\(\frac{G}{F_S}\)^2F_\phi+h.c.\]\\
        &+\kappa_{\phi\bar{\phi}}\[\frac{i}{2}|F_\phi|^2\partial_\mu\(\frac{G}{F_S}\)\sigma^\mu\frac{\bar{G}}{\bar{F}_S}+\frac{i}{2}\frac{G\sigma^\mu\bar{G}}{|F_S|^2}\bar{F}_\phi\partial_\mu F_\phi-F_\phi\frac{G}{F_S}\sigma^\mu\bar{\sigma}^\nu\partial_\nu\(\frac{G}{F_S}\)\partial_\mu A\right.\\
        &\left.-\(\frac{G}{F_S}\)^2\partial_\mu F_\phi\partial^\mu A+i\partial^\mu\(\frac{G}{F_S}\)\sigma^\nu\frac{\bar{G}}{\bar{F}_S}\partial_\mu A\partial_\nu A-\frac{i}{2}\partial_\mu\(\frac{G}{F_S}\)\sigma^\mu\frac{\bar{G}}{\bar{F}_S}\(\partial A\)^2+h.c.\]\\
        &-\(\frac{i}{2}g''F_\phi+\frac{i}{2}f'F_S+h.c.\)\frac{G\sigma^\mu\bar{G}}{|F_S|^2}\partial_\mu A-\frac{1}{4|F_S|^2}\(1+\frac{\kappa_{\phi\bar{\phi}}|F_\phi|^2}{2|F_S|^2}\)\bar{G}^2\Box G^2\\
        &+\(f''F_S+g'''F_\phi+h.c.\)\frac{G^2\bar{G}^2}{16|F_S|^4}|F_\phi|^2+\(\kappa_{\phi\bar{\phi}\phi\phi}+\kappa_{\phi\bar{\phi}\bar{\phi}\bar{\phi}}+2\kappa_{\phi\bar{\phi}\phi\bar{\phi}}\)\frac{G^2\bar{G}^2}{16|F_S|^4}|F_\phi|^4\\
        &-\frac{1}{2}\[\(\frac{\kappa_{\phi\bar{\phi}\phi}+\kappa_{\phi\bar{\phi}\bar{\phi}}}{2}+g''\)\(\frac{G}{F_S}\)^2\(\partial A\)^2+h.c.\]+\frac{\kappa_{\phi\bar{\phi}}}{16}\[F_\phi^2\(\frac{G}{F_S}\)^2\Box \(\frac{G}{F_S}\)^2+h.c.\].
    \end{aligned}
\end{equation}

As anticipated, we have that also $F_\phi$, now unconstrained, enters the lagrangian as an auxiliary field which needs to be integrated out together with $F_S$. The resulting on-shell lagrangian relevant for the quadratic 
limit is then
\begin{equation}
    \begin{aligned}
        \mathcal{L}=&-\[|f|^2+\frac{|g'|^2}{\kappa_{\phi\bar{\phi}}}\]+\kappa_{\phi\bar{\phi}}\partial_\mu A\partial^\mu A+Z\(\frac{i}{2}\partial_\mu G\sigma^\mu\bar{G}+h.c.\)+\(mG^2+h.c.\)+c\,G^2\bar{G}^2\\
        &+Z\(i\frac{f'}{2f}+h.c.\) G\sigma^\mu\bar{G}\partial_\mu A+\(d\,G^2+h.c.\)\(\partial A\)^2-\frac{1}{4|f^2|}\(1+\frac{|g'|^2}{2\kappa_{\phi\bar{\phi}}|f|^2}\)\bar{G}^2\Box G^2\\
        &+\[\frac{\bar{g}'}{\bar{f}^2}\partial_\mu G\sigma^\mu\bar{\sigma}^\nu G\partial_\nu A+\frac{\kappa_{\phi\bar{\phi}}}{|f|^2}i\partial^\mu G\sigma^\nu\bar{G}\partial_\mu A\partial_\nu A-\frac{\kappa_{\phi\bar{\phi}}}{2|f|^2}i\partial_\mu G\sigma^\mu\bar{G}\(\partial A\)^2+h.c.\] \\
        &+\frac{1}{16\kappa_{\phi\bar{\phi}}}\(\frac{\bar{g}'^2}{\bar{f}^4}G^2\Box G^2+h.c.\)\ ,
    \end{aligned}
\label{eqn:gc_lag_ql}
\end{equation}

with 
\begin{equation}
    Z\equiv 1+\frac{|g'|^2}{\kappa_{\phi\bar{\phi}}|f|^2}.
\label{eqn:gc_g_wfr}
\end{equation}
We do not specify further the other indicative coefficients, i.e. $m$, $c$ and $d$, since they are not going to be relevant for our discussion. Up to quartic pure goldstino terms, the lagrangian (\ref{eqn:gc_lag_ql}) corresponds to the quadratic limit lagrangian in this framework with improved constraints. The crucial difference with respect to the orthogonal constraint case described in Section \ref{section:goldstino-actions} is that here the scalar potential is restored to be the standard supersymmetric one
\begin{equation}
    V=|F_S|^2 +K_{\phi\bar{\phi}}|F_{\phi}|^2 = |\partial_S W|^2 + \frac{|\partial_{\phi} W|^2}{K_{\phi\bar{\phi}}}=|f|^2+\frac{|g'|^2}{\kappa_{\phi\bar{\phi}}},
\label{eqn:gc_new_sp}
\end{equation}
since now both auxiliary fields are determined algebraically by the usual rules of supersymmetric lagrangians. The counterpart of this change is that the goldstino kinetic term also gets shifted, acquiring the normalization factor $Z$ given in (\ref{eqn:gc_g_wfr}). 

Instead, the lagrangian in the four-field basis is obtained via the field redefinitions
\begin{align}
    \delta G^\alpha=&\frac{\kappa_{\phi\bar{\phi}}}{2Z|f|^2}G^\alpha\(\partial A\)^2, & \delta G^\alpha=& -\frac{i}{Z}\frac{g'}{f^2}\(\bar{G}\bar{\sigma}^\mu\)^\alpha\partial_\mu A,
\end{align}
and become
\begin{equation}
    \begin{aligned}
        \mathcal{L}=&-\[|f|^2+\frac{|g'|^2}{\kappa_{\phi\bar{\phi}}}\]+\kappa_{\phi\bar{\phi}}\partial_\mu A\partial^\mu A+Z\(\frac{i}{2}\partial_\mu G\sigma^\mu\bar{G}+h.c.\)\\
        &+\(mG^2+h.c.\)+\lambda G\sigma^\mu\bar{G}\partial_\mu A+c\,G^2\bar{G}^2+\(\tilde{d}\,G^2+h.c.\)\partial_\mu A\partial^\mu A\\
        &+\(\frac{\kappa_{\phi\bar{\phi}}}{|f|^2+\frac{|g'|^2}{\kappa_{\phi\bar{\phi}}}}\)\(i\partial^\mu G\sigma^\nu\bar{G}\partial_\mu A\partial_\nu A+h.c.\)-\frac{1}{4|f|^2}\(1+\frac{|g'|^2}{2\kappa_{\phi\bar{\phi}}|f|^2}\)\bar{G}^2\Box G^2 \\
        &+\frac{1}{16\kappa_{\phi\bar{\phi}}}\(\frac{\bar{g}'^2}{\bar{f}^4}G^2\Box G^2+h.c.\)\ ,
    \end{aligned}
\label{eqn:gc_lag_tot}
\end{equation}

\noindent where, again, the coefficients $\lambda$ and $\tilde{d}$ are not specified further because they are not relevant for our analysis.
\subsection{Sound speed and (trivial) positivity bounds}

The lagrangian (\ref{eqn:gc_lag_tot}) immediately gives confirmation to our starting claim: the coefficients of the operators $\(i\partial^\mu G\sigma^\nu\bar{G}\partial_\mu A\partial_\nu A+h.c.\)$ and $\bar{G}^2\Box G^2$ now identically satisfy the positivity bounds required to ensure causality. Also the improved bound on non-elastic scattering becomes trivial.

Again, looking at the goldstino sound speed in a time-dependent scalar field background $A=A(t)$, we understand that the theory is actually well-defined in the whole field space, not only in the vacuum. Indeed, following the same steps of Section \ref{section:goldstino-actions}\footnote{The operator $G\sigma^\mu\bar{G}\partial_\mu A$ can again be seen not to contribute to the sound speed. The argument is the one presented in footnote \ref{footnotef'}.}, we see that thanks to the modification of the goldstino kinetic term normalization (\ref{eqn:gc_g_wfr}), connected to that of the scalar potential (\ref{eqn:gc_new_sp}), the sound speed becomes
\begin{equation}
    c_s^2=1-\frac{4{\dot A}^2|f^2|\kappa_{\phi\bar{\phi}}}{\(|f|^2+\frac{|g'|^2}{\kappa_{\phi\bar{\phi}}}+\kappa_{\phi\bar{\phi}}{\dot A}^2\)^2} \ .
\end{equation}
This sound speed exactly matches the SUGRA one (\ref{eq:orthog8}) and it is subluminal along the whole  scalar field evolution.

Therefore, in this improved framework it is not necessary to impose any extra condition on the theory parameters: the two single constraints (\ref{eqn:gc_sc}) and (\ref{eqn:gc_ferm}) give rise to a theory with the same field content as the orthogonal constraint case, but which is now causal on the whole field space. This indicates that the theory can be completed in the UV into what would then be a minimal inflationary model. 

This result supports our understanding of the (potential) inconsistencies associated to the orthogonal constraint (\ref{eqn:o_const}). The improvement we have found in the generalized constraints setup described in this section lies in the fact that the two single constraints (\ref{eqn:gc_sc}) and (\ref{eqn:gc_ferm}) effectively realize the decoupling of a heavy scalar and of a heavy fermion, respectively: being those physical degrees of freedom, this procedure is well-understood and legitimate, and it does not lead to any causal inconsistency in the IR theory. On the contrary, the orthogonal constraint treats on the same footing the auxiliary field by also enforcing the third constraint (\ref{eqn:gc_aux}), although "decoupling" an auxiliary field does not have a proper physical meaning unless one adds higher-derivative operators in the UV and this can lead to acausality behaviours in the IR theory, as described in Section \ref{section:goldstino-actions}. In the next section, we will see that adding higher-derivative operators in the UV does change in fact this conclusion. 

\section{Lagrangians with orthogonal constraints and no causality condition }\label{section:orthogonal-nocondition}

All the above superfield constraints which resulted in non-trivial positivity bounds share the feature that they "eliminate" an auxiliary field in terms of the other degrees of freedom. In the language of Section~\ref{section:twoconstraints}, their solution verifies $S\bar SD^2Q_L=0$, where $Q_L$ is the constrained superfield. This led us to suspect that eliminating the auxiliary field by a superfield constraint has no simple interpretation in a microscopic two-derivative SUSY/SUGRA theory, and is responsible for the breakdown of positivity.

One may wonder how this claim is articulated with the fact that auxiliary fields, when present, can always be integrated out and expressed in terms of other degrees of freedom, without leading to any breakdown of positivity. One tentative answer could be that integrating out auxiliary fields yields a theory where SUSY is only realized on-shell, while the constraint $S\bar SD^2Q_L=0$, which removes the auxiliary field, allows to write actions with off-shell SUSY. Instead, we will explain in this section that {\it i)} off-shell nonlinear SUSY, i.e. the explicit presence of the nilpotent field $S$, allows one to integrate out auxiliary fields while retaining off-shell SUSY by working in superspace, that {\it ii)} this process generates the constraint $S\bar SD^2Q_L=0$ {\it and} higher-derivative interactions in addition to the two-derivative lagrangian, and that {\it iii)} these additional terms make the positivity bounds trivially satisfied. This shows that the non-trivial causality constraints signal that the use of constraints which remove auxiliary fields clash with the use of two-derivative actions for the associated constrained superfields.

We believe that the use of improved constraints, as in Section~\ref{section:twoconstraints}, and the approach of the present section are equivalent. Indeed, one assumes the same microscopic action and superfield content in both cases, and one subsequently only solves equations of motion associated to an auxiliary field, either in component or in superspace. The lagrangians obtained in both approaches should be related by field redefinitions, but we have not tried to perform this identification systematically.

\subsection{Integrating out an auxiliary field in superspace}\label{Fint}

Let us consider a chiral superfield $Q_L$, and show how one can integrate out its auxiliary field without giving up off-shell supersymmetry. This can be achieved due to the power of off-shell nonlinear SUSY associated to $S$, thanks to which we can define a chiral multiplet $ Q_L^{(a)}$,
\beq
 Q_L^{(a)}\equiv \frac{S}{\bar D^2\bar S}\bar D^2\(\frac{\bar S D^2Q_L}{D^2 S}\) \ ,
\label{defSubFields}
\eeq
which is by construction nilpotent, i.e. $Q_L^{(a)}{}^2=0$, and such that\footnote{All these relations follow from $S^2=0$ (and its consequence $SD_\alpha S=0$). For instance, when computing $S\bar SD^2Q^{(a)}_L$, the available $D^2$ and $\bar D^2$ must fully act on the $S$ and $\bar S$ factors in $Q^{(a)}_L$, respectively:
\bes
S \bar S D^2 Q_L^{(a)}=S \bar S D^2\(\frac{S}{\bar D^2\bar S}\bar D^2\(\frac{\bar S D^2Q_L}{D^2 S}\)\)=S \bar S \frac{D^2 S}{\bar D^2\bar S}\bar D^2\(\frac{\bar S D^2Q_L}{D^2 S}\)=S \bar S \frac{D^2 S}{\bar D^2\bar S}\frac{\bar D^2\bar S D^2Q_L}{D^2 S}=S \bar SD^2Q_L \ .
\ees} 
\beq
\quad S \bar S  Q_L^{(a)}=S \bar S D_\alpha Q_L^{(a)}=S \bar S (D^2 Q_L^{(a)}-D^2Q_L)=0 \ .
\eeq
The first two constraints imply that the would-be dynamical scalar and fermion of $ Q_L^{(a)}$ are expressed in terms of its auxiliary field. The last one implies that, for any $Q_L$, one can write
\beq
Q_L=Q_L'+ Q_L^{(a)} \ , \quad \text{with } \quad S \bar S  Q_L^{(a)}=S \bar S D_\alpha Q_L^{(a)}=S \bar S D^2Q_L'=0 \ ,
\eeq
i.e. $Q_L$ is the sum of a superfield which contains only a dynamical complex scalar and a dynamical fermion ($Q_L'$) and a superfield which contains only an auxiliary field ($ Q_L^{(a)}$). In addition, if $Q_L$ verifies constraints such as $S \bar S Q_L=0$ or $S \bar S D_\alpha Q_L=0$, so does $Q_L'$.

$ Q_L^{(a)}$ being fully auxiliary, it can be integrated out, and this can be explicitly performed in superspace. Let us take the action for $Q_L$ to be a regular two derivative action, whose lagrangian reads
\beq
\bead
\cL&=\int d^4\theta K(S,\bar S,Q_L,\bar Q_L) +\[\int d^2\theta W(S,Q_L)+h.c.\]\\
&=\int d^4\theta K(S,\bar S,Q_L'+ Q_L^{(a)},\bar Q_L'+\bar Q_L^{(a)})+\[\int d^2\theta W(S,Q_L'+ Q_L^{(a)})+h.c.\]  \ .
\eead
\eeq
One can integrate out the auxiliary superfield $ Q_L^{(a)}$, i.e. solve its equation of motion and express it in terms of $Q_L'$, in order to get an action for $Q_L'$ only, of which we remind that it verifies by construction $S \bar S D^2Q_L'=0$. 

When we perform this integration, we need to account for the fact that the superfields are constrained (beyond being chiral). Hence, we add to the action two unconstrained Lagrange multiplier superfields $M_1$ and $M_2^\alpha$,
\beq
\cL \to \cL +\int d^4\theta S \bar S \(M_1 Q_L^{(a)}+M_2^\alpha D_\alpha  Q_L^{(a)}\)+h.c. \ ,
\eeq
so that the equation of motion for $Q_L^{(a)}$ reads
\beq
\label{eomPhia}
\bar D^2 \partial_{Q_L} K=4\partial_{Q_L}W-\bar D^2\(S \bar S M_1+D_\alpha\[S \bar S M_2^\alpha\]\) \ .
\eeq
The equations of motion of $M_{1,2}$ enforce the constraints on $ Q_L^{(a)}$. Upon acting with $S \bar S$ on \eqref{eomPhia}, one finds
\beq
\label{soleomPhia}
 Q_L^{(a)}=\bar D^2\(\frac{\abs{S}^2\[4\partial_{\bar Q_L}\bar W(0,\bar Q_L')-\partial^2_{S,\bar Q_L} K(0,0,Q_L',\bar Q_L')D^2S\]}{\partial^2_{Q_L,\bar Q_L} K(0,0, Q_L',\bar Q_L')\abs{D^2S}^2}\) \ .
\eeq
This solution is such that the Lagrange multipliers disappear from the action. In order to derive it, we used the fact that the constraints on $ Q_L^{(a)}$ imply the following identity,
\beq
 Q_L^{(a)}=\bar D^2\(\frac{\abs{S}^2D^2 Q_L^{(a)}}{\abs{D^2 S}^2}\) \ ,
\label{solPhia}
\eeq
and that\footnote{The relation $S  Q_L^{(a)}=0$ follows from \eqref{solPhia} and is not an extra constraint. As for \eqref{solPhia}, it follows from $S \bar S  Q_L^{(a)}=S \bar S D_\alpha Q_L^{(a)}=0$. Indeed, acting with $\bar D^2$ on the first equality removes $\bar S$ and leaves out the chiral constraint $S  Q_L^{(a)}=0$, which can be solved with a subsequent action of $D^2$,
\bes
 Q_L^{(a)}=-\frac{S D^2 Q_L^{(a)}+2D^\alpha S D_\alpha Q_L^{(a)}}{D^2 S} \ .
\ees
We can then multiply by $\bar S$ and act with $\bar D^2$ to obtain
\bes
 Q_L^{(a)}=-\frac{1}{\bar D^2\bar S}\bar D^2\(\bar S \frac{S D^2 Q_L^{(a)}+2D^\alpha S D_\alpha Q_L^{(a)}}{D^2S}\)=\frac{1}{\bar D^2\bar S}\bar D^2\(\bar S\frac{S D^2 Q_L^{(a)}}{D^2S}\) \ ,
\ees
corresponding to the expression in \eqref{solPhia}. In the last step, we used the result of acting with $D^\alpha$ on $S \bar S D_\alpha Q_L^{(a)}=0$, namely $\bar S \[D^\alpha S D_\alpha Q_L^{(a)}+S D^2 Q_L^{(a)}\]=0$.} $S^2=S  Q_L^{(a)}=S \bar S D_\alpha  Q_L^{(a)}=0$. 

Eventually, inserting \eqref{soleomPhia} in the lagrangian results in the following action (see Appendix \ref{appendix:IRlagNoF} for details),
\beq
\label{lagAfterIntPhia}
\bead
\cL_\text{eff}=&\int d^4\theta K(S,\bar S, Q_L',\bar  Q_L')+\[\int d^2\theta W\(S, Q_L'\) +h.c.\]\\
&+\int d^4\theta\frac{1}{\partial^2_{Q_L,\bar Q_L}K(0,0, Q_L',\bar Q_L')}\abs{\frac{S}{\bar D^2\bar S}}^2\Bigg(-\abs{4\partial_{Q_L}W(0,\bar Q_L')-\partial^2_{Q_L,\bar S} K(0,0, Q_L',\bar Q_L')\bar D^2\bar S}^2\\
&+\left\{\partial^3_{Q_L,\bar Q_L^2}K(0,0, Q_L',\bar Q_L')\[\bar D\bar Q_L'\]^2\[4\partial_{\bar Q_L}\bar W(0,\bar Q_L')-\partial^2_{S,\bar Q_L} K(0,0, Q_L',\bar Q_L')D^2S\]+h.c.\right\}\Bigg) \ .
\eead
\eeq
The first line is simply the two-derivative action that we started from, now evaluated on $ Q_L'$ which verifies $S\bar S D^2 Q_L'=0$. As anticipated, it is supplemented by higher-derivative interactions. We now proceed to show that these extra terms yield exactly what is required to trivialize all the causality conditions.

\subsection{Application to models of real scalars}\label{realScalars}

We first consider the models explored in Section~\ref{section:orthogonal}. Our goal here is to start with the same dynamical degrees of freedom and the same two-derivative action, but instead of constraining the auxiliary field, we integrate it out following the construction of the previous section. Therefore, we take $Q_L$ to verify $S \bar S D_\alpha  Q_L=S \bar S (Q_L-\bar Q_L)=0$, which only leaves a dynamical real scalar $A$ behind (together with the auxiliary field). As mentioned above, these constraints also apply to $ Q_L'$, which we identify with $\Phi$ of Section~\ref{section:orthogonal}, as they verify the same constraints. We also use $K=h\(\frac{\Phi+\bar\Phi}{2}\)\(\frac{\Phi-\bar \Phi}{2i}\)^2+ S \bar S$ and $W=f(\Phi)S +g(\Phi)$, as in Section~\ref{section:orthogonal}. 

The constraints on $Q_L'=\Phi$ as well as the form of the K\"ahler potential drastically simplify the terms which arise from the second and third lines of \eqref{lagAfterIntPhia} and which should be added to \eqref{eqn:o_LKW}. They simply read
\beq
\label{simplerExtraTermRealScalar}
\delta \cL =-16\int d^4\theta\frac{1}{\partial^2_{\Phi,\bar \Phi}K(0,0, \Phi,\bar \Phi)}\abs{\frac{S}{\bar D^2\bar S}\partial_{\Phi}W(0,\bar \Phi)}^2 \ .
\eeq
We then insert the expressions of $K,W$ and of $ \Phi$ in terms of $A$ presented in Section~\ref{section:orthogonal}. Dropping the terms irrelevant in both the quadratic and four-field limits, one eventually finds
\beq
\label{deltaLRealScalar}
\bead
\delta \cL=&-\frac{2\abs{g'}^2}{h}\[1+\abs{\frac{1}{F_S}}^2\Bigg(\frac{i}{2}G\sigma^\mu\partial_\mu\bar G-\frac{i}{2}\partial_\mu G\sigma^\mu\bar G+\frac{\abs{\partial_\mu s}^2}{2}+3\frac{\bar s \Box s + s\Box \bar s}{4}\Bigg) \]\\
&+\frac{i\abs{g'}^2}{\abs{F_S}^2h}G\sigma^\mu \bar G\(\frac{\partial_\mu\bar F_S}{\bar F_S}-h.c.\)\ ,
\eead
\eeq 
where the $A$-dependence of the scalar functions is left implicit, and we remind that $s= \frac{G^2}{2F_S}$. After solving for\footnote{Note that the equation of motion of $F_S$ is modified by \eqref{deltaLRealScalar} with respect to \eqref{eqn:o_LKW}, but not at the bosonic level, while the additional two-fermion terms are all proportional to the free equation of motion $\sigma^\mu\partial_\mu G$. In the four-field limit, that allows us to simply use the bosonic value of $F_S$ in \eqref{deltaLRealScalar}.} $F_S$, integrating by part and dropping further irrelevant terms, we eventually find
\beq
\delta \cL=-\frac{2\abs{g'}^2}{h}-\frac{2\abs{g'}^2}{h\abs{f}^2}\(\[\frac{i}{2}G\sigma^\mu\partial_\mu\bar G+h.c.\]-\(i\frac{f'}{2f}+h.c.\)G\sigma^\mu\bar G\partial H+ \frac{1}{4\abs{f}^2}G^2 \Box \bar G^2\) \ ,
\eeq
and the resulting lagrangian can be matched to that of \eqref{eqn:o_L_tot}, upon making the fermion kinetic term canonical,
\beq
G\to \(1+\frac{2\abs{g'}^2}{h\abs{f}^2}\)^{-1/2}G \ ,
\eeq
and redefining
\beq
f\to \tilde f\equiv f\(1+\frac{2\abs{g'}^2}{hf^2}\)^{\frac{1}{2}} \ .
\eeq
All other couplings are left untouched\footnote{The fact that $\delta \cL$ merely amounts to a shift of parameters in \eqref{eqn:o_L_tot} is a consequence of nonlinear supersymmetry, which imposes that the operator coefficients are all determined by a small set of functions. For instance, the vacuum energy is fixed given the goldstino decay constant, which is (up to a numerical coefficient) the coefficient of the operator $G^2\Box \bar G^2$ when the goldstino is canonically normalized. The same applies for the coupling $\partial^\mu G \sigma^\nu \bar G \partial_\mu A \partial_\nu A$ when $A$ is canonically normalized.}. Note that $\delta \cL$ adds a term to the scalar potential of \eqref{eqn:o_L_final}, such that the scalar potential accounting for all auxiliary fields is recovered.

Having matched to \eqref{eqn:o_L_tot}, we can use the results of Section~\ref{section:orthogonal}, leading to the causality/positivity bound
\beq
h\tilde f^2>2 g'^2 \iff  h f^2 >0 \ ,
\eeq
which is trivially satisfied. Therefore, the addition of $\delta\cL$ ensured that there is no non-trivial causality constraint, despite the fact that $\Phi$ is constrained not to contain an auxiliary field.

\subsection{Application to models of complex scalars}\label{complexScalars}

We now turn to the models of Section~\ref{section:complex}. Those contain a complex scalar $H$ but no dynamical fermion, hence we take $Q_L$ to verify $S \bar S D_\alpha Q_L=0$. As explained above, this also implies that $S \bar S D_\alpha  Q_L'=0$, hence $S \bar Q_L'=$ chiral, and we can identify $Q_L'=H$. Following Section~\ref{section:complex}, we use $K=\xi(H,\bar H) S \bar S+\kappa(H,\bar H)$ and $W=f(H)S+g(H)$.

As above, the constraints on $H$ and the form of the K\"ahler potential allow us to use \eqref{simplerExtraTermRealScalar}, where we insert the expressions $K,W$ and of $H$ presented in Section~\ref{section:complex}. Dropping all terms irrelevant for both the quadratic and four-field limits, one eventually finds
\beq
\label{complexOF}
\bead
\delta \cL=&-\frac{\abs{g'}^2}{\kappa_{H\bar H}}\[1+\abs{\frac{1}{F_S}}^2\Bigg(\frac{i}{2}G\sigma^\mu\partial_\mu\bar G-\frac{i}{2}\partial_\mu G\sigma^\mu\bar G+\frac{\abs{\partial s}^2}{2}+3\frac{\bar s \Box s + s \Box \bar s}{4}\Bigg) \]\\
&-\frac{i}{2\abs{F_S}^2\kappa_{H\bar H}}G\sigma^\mu \bar G\(\abs{g'}^2\[\frac{\kappa_{H^2\bar H}}{\kappa_{H \bar H}}\partial_\mu H-\frac{\partial_\mu\bar F_S}{\bar F_S}\]-\bar g'g''\partial_\mu H-h.c.\)\ ,
\eead
\eeq 
where the $H$-dependence of the scalar functions is left implicit. Fixing $\xi=1$, solving for $F_S$, integrating by part and dropping further irrelevant terms, we eventually find
\beq
\bead
\delta \cL=-\frac{\abs{g'}^2}{\kappa_{H\bar H}}&-\frac{1}{\kappa_{H\bar H}}\abs{\frac{g'}{f}}^2\(\[\frac{i}{2}G\sigma^\mu\partial_\mu\bar G+h.c.\]+ \frac{1}{4\abs{f}^2}G^2\Box \bar G^2\)\\
&-\frac{i}{2\kappa_{H\bar H}\abs{f}^2}G\sigma^\mu \bar G\(\partial_\mu H\[\abs{g'}^2\(\frac{\kappa_{H^2\bar H}}{\kappa_{H\bar H}}-\frac{f'}{f}\)-\bar g'g''\]-h.c.\) \ .
\eead
\eeq 
The resulting lagrangian can be matched to that of \eqref{eqn:cs_L_on}, upon redefining
\beq
G \to \tilde G\equiv\(1+\frac{1}{\kappa_{H\bar H}}\abs{\frac{g'}{f}}^2\)^{-\frac{1}{2}}G \ , \quad f \to \tilde f\equiv \(1+\frac{1}{\kappa_{H\bar H}}\abs{\frac{g'}{f}}^2\)^{\frac{1}{2}}f \ ,
\eeq
all other fields and couplings being left untouched. As in the previous section, $\delta\cL$ adds a term to the scalar potential of \eqref{eqn:cs_L_tot}, such that the scalar potential accounting for all auxiliary fields is recovered.

Having matched to \eqref{eqn:cs_L_on}, we can use the results of Section~\ref{section:complex}, leading to the causality/positivity bound
\beq
\tilde\kappa_{H\bar H}\abs{\tilde f}^2>\abs{\tilde g'}^2 \iff \kappa_{H\bar H}\abs{f}^2>0 \ ,
\eeq
which is trivially satisfied. Therefore, the addition of $\delta\cL$ ensured that there is no non-trivial causality constraint, despite the fact that $H$ is constrained not to contain an auxiliary field.

\section{Minimal models of inflation with no causality constraints }\label{section:inflation}

The alternative models put forward in the last Sections 
\ref{section:twoconstraints} and \ref{section:orthogonal-nocondition} can accommodate a minimal physical spectrum from the viewpoint of an inflationary model in supergravity: the graviton, a massive gravitino and a real scalar, the inflaton. They share this feature with models using the orthogonal constraint (and two-derivative actions). They have the advantage however to remove any causality condition on the theory parameter space. As stressed previously, we believe the two alternative approaches
in Section~\ref{section:twoconstraints} and Section~\ref{section:orthogonal-nocondition} are equivalent and lead to the same physical observables. In particular, the scalar potential in both cases is the usual supersymmetric one. For supergravity models, it is therefore given by the standard SUGRA formulae \cite{Cremmer:1982en}.    

Building minimal models of inflation along these lines is straightforward. Let us consider models for which the two-derivative part of the action is
\begin{equation}
K = - \frac{1}{2} (\Phi - {\bar \Phi})^2 + {\bar S} S \quad , \quad W = f(\Phi) S + g(\Phi) \ , \label{eq:mmi1}    
\end{equation}
where $S$ is nilpotent and $\Phi$ contains as only physical degree of freedom the inflaton $\varphi=\Re(\Phi)|$. The scalar potential is
\begin{equation}
V = |f(\varphi)|^2 + |g'(\varphi)|^2 - 3 |g(\varphi)|^2    \ . \label{eq:mmi2}  
\end{equation}
We restrict for simplicity to the class of models put forward in \cite{DallAgata:2014qsj}, defined by 
$f = \sqrt{3} g$, for which the scalar potential reduces
to 
\begin{equation}
V = |g'(\varphi)|^2   \ . \label{eq:mmi3}    
\end{equation}
Let us give two simple examples of inflationary models.
\paragraph{Starobinsky model} One chooses
\begin{equation}
g = M^2 \left (\Phi + \frac{1}{a} e^{-a \Phi} \right) + 
g_0   \ , \label{eq:mmi4}    
\end{equation}
where $M$ is a mass scale which will fix the energy scale during inflation, whereas $g_0$ will determine supersymmetry breaking in the vacuum. One gets the scalar
potential
\begin{equation}
V (\varphi) = M^4 [ 1-e^{-a \varphi}]^2   \ , \label{eq:mmi5}  
\end{equation}
which is the usual Starobinsky scalar potential of the inflaton $\varphi$. 

\paragraph{Chaotic inflation} One chooses
\begin{equation}
g = \frac{m}{2\sqrt{2}} \Phi^2 + 
g_0   \ , \label{eq:mmi6}    
\end{equation}
where $m$ is the inflaton mass. 
The scalar potential becomes now the simplest example of chaotic inflation 
\begin{equation}
V (\varphi) = \frac{m^2 \varphi^2}{2}   \ . \label{eq:mmi7}  
\end{equation}
As explained in \cite{DallAgata:2014qsj}, the advantage of such models is the decoupling of the energy scale of inflation and that of supersymmetry breaking in the vacuum. Indeed, the latter is governed by the independent parameter $g_0$, which can be chosen such that the supersymmetry breaking scale is much smaller than the inflationary scale. 

Finally, let us stress that our alternative inflationary models have the same minimal particle content as models based on the orthogonal constraint  \cite{Ferrara:2015tyn,Carrasco:2015iij}, but that the scalar potential, therefore the inflationary dynamics, are different. gravitino/goldstino interactions to the inflaton also differ.

\section{Conclusions and perspectives}\label{section:conclusions}

In this paper we investigated causality constraints on supergravity and supersymmetry models with nonlinear supersymmetry. By using the equivalence theorem, the causality constraints from the sound speed of the gravitino in SUGRA models translate in the $M_P\to\infty$ limit into causality/positivity constraints in low-energy goldstino lagrangians. We found nontrivial causality conditions on the theory parameter space of models with superfield constraints which "eliminate" auxiliary fields. Example in this class are the orthogonal constraint which eliminates a real (or pseudoreal) scalar, a fermion and an auxiliary field,  and the constraint eliminating a fermion and an auxiliary field \cite{Komargodski:2009rz}. We found that in general the SUGRA causality condition has gravitational corrections, which should imply that small gravitational (of order $1/M_P^2$) violations of causality/positivity in the low-energy goldstino actions are allowed. This is in agreement with causality arguments discussed in the literature \cite{Alberte:2020jsk}. We stress that the causality bounds related to the gravitino/goldstino sound speed are valid not only in the ground state, but for all values of the scalar fields which solve the equations of motion, such as time-dependent inflationary dynamics in supergravity or solutions in the low-energy theory. While, as we demonstrated explicitly, they agree in the ground state with positivity arguments on $2 \to 2$ particles scattering, they hold more generally and are therefore stronger.

Unless the low-energy functions, depending on theory parameters and scalar fields, are restricted to satisfy the causality constraints {\it identically}, we interpret the constraints as an obstruction to a microscopic two-derivative UV completion of such models. Since the decoupling of scalars and/or fermions should not be problematic, we argued that the problem comes from constraints which eliminate auxiliary fields. We substantiated this claim in two ways. First, by imposing alternative superfield constraints which eliminate the same physical fields, but not  the auxiliary field, and showing that causality is automatically guaranteed, i.e. that the causality constraints are made trivial. Secondly,  by explicitly integrating auxiliary fields in complete theories coupled to nonlinear SUSY, generating along the way higher-derivative operators which again trivialize the causality constraints.  Let us emphasize again that when the causality constraints are satisfied identically, there should be no obstruction for a microscopic completion. This is the case for example for the orthogonal constraint with a constant gravitino mass, corresponding to $g'=0$. 

Using our constructions, we provided new SUGRA models for inflation, with a minimal spectrum (a graviton, a massive gravitino and a real inflaton) and no causality constraints. Unlike the case of the orthogonal constraint, the scalar potential is provided by the standard SUGRA formulae \cite{Cremmer:1982en}.  

It would be interesting to provide a general analysis of causality constraints in nonlinearly realized supersymmetric models in supergravity, in the spirit of the swampland program. We notice that important steps from the stability viewpoint in such constructions were taken  recently in string models with brane supersymmetry breaking \cite{bsb} and nonlinear supersymmetry \cite{dm}, see e.g. \cite{basile}. 

\vskip 1cm
\section*{Acknowledgments} 

We would like to thank Brando Bellazzini for useful discussions. Q.B. is supported by the Deutsche Forschungsgemeinschaft under Germany’s Excellence Strategy EXC 2121 “Quantum Universe” - 390833306, and by the Deutsche Forschungsgemeinschaft under the grant 491245950.

\begin{appendices}

\section{Explicit solution of constraints (\ref{eqn:gc_sc})
and (\ref{eqn:gc_ferm})} 
\renewcommand{\theequation}{A.\arabic{equation}}\label{appendix:solutionsConstraints}
\setcounter{equation}{0}

We start with the solution of the generalized constraint (\ref{eqn:gc_sc}), imposed alone. This fixes the imaginary part $B$ of the scalar field $\phi$ as the function of the goldstino $G$, the fermion $\psi$ and the auxiliary fields $F_S$, $F_{\phi}$. Starting from the generalized constraint solution (\ref{eqn:gen_const_sol}), the equation for $B$ results to be
\begin{equation}
\begin{aligned}
	B=&\frac{i}{2}\left(\frac{\bar G \bar{\psi}}{\bar F_S}-\frac{G\psi}{F_S}\right)+\frac{i}{4}\left[\left(\frac{G}{F_S}\right)^2 F_\phi-\left(\frac{\bar G}{\bar F_S}\right)^2\bar F_\phi\right]-\frac{G\sigma^\mu\bar G}{2\abs{F_S}^2}\partial_\mu A\\
	&+\frac{\psi\sigma^\mu\partial_\mu \bar G}{2\abs{F_S}^2}\frac{G^2}{2F_S}-\frac{G\sigma^\mu
	\bar{\psi}}{2\abs{F_S}^2}\partial_\mu\left(\frac{\bar G^2}{2\bar F_S}\right)-\frac{G\sigma^\mu\partial_\mu \bar{\psi}}{2\abs{F_S}^2}\frac{\bar G^2}{2\bar F_S}\\
	&+\frac{i}{2\bar F_S}\left(\frac{G}{F_S}\right)^2\partial_\mu\left(\frac{\bar G^2}{2\bar F_S}\right)\partial^\mu A+i\frac{G^2\bar G^2}{8\abs{F_S}^4}\Box A\\
	&+i\frac{G\sigma^\mu\partial_\mu\bar G}{\abs{F_S}^2}B+\frac{i}{2}\frac{G\sigma^\mu\bar G}{\abs{F_S}^2}\partial_\mu B+\frac{1}{2\bar F_S}\left(\frac{G}{F_S}\right)^2\Box\left(\frac{\bar G^2}{2\bar F_S}\right)B\\
	&+\frac{1}{2\bar F_S}\left(\frac{G}{F_S}\right)^2\partial_\mu\left(\frac{\bar G^2}{2\bar F_S}\right)\partial^\mu B+\frac{G^2\bar G^2}{8\abs{F_S}^4}\Box B \ .
\end{aligned}
\label{eqn:gc_eq_B}
\end{equation}
This equation can be solved iteratively via an expansion of operators with increasing number of fermions, so that one finds
\begin{equation}
	B=B^{(2)}+B^{(4)}+B^{(6)}+B^{(8)} \ ,
\end{equation}

where the number in the superscripts denotes the number of goldstino fields. The explicit expressions for $B^{(2)}$ and $B^{(4)}$ are, respectively,
\begin{align}
	&B^{(2)}= \frac{i}{2}\left(\frac{\bar G\bar \psi}{\bar F_S}-\frac{G\psi}{F_S}\right)+\frac{i}{4}\left[\left(\frac{G}{F_S}\right)^2 F_\phi-\left(\frac{\bar G}{\bar F_S}\right)^2\bar F_\phi\right]-\frac{G\sigma^\mu\bar G}{2\abs{F_S}^2}\partial_\mu A \ ,
\label{eqn:gc_B2}\\\notag\\
	&\begin{aligned}
	    B^{(4)}=&-\frac{G\sigma^\mu\partial_\mu\bar G}{\abs{F_S}^2}\frac{\bar G\bar \psi}{2\bar F_S}+\frac{G\sigma^\mu\bar G}{4\abs{F_S}^2}\left[\psi\partial_\mu\left(\frac{G}{F_S}\right)-\bar \psi\partial_\mu\left(\frac{\bar G}{\bar F_S}\right)\right]\\
	&-\left[\frac{G^2}{8F_S}\frac{\partial_\mu\psi\sigma^\mu\bar G}{\abs{F_S}^2}+\frac{\bar G^2}{8\bar F_S}\frac{G\sigma^\mu\partial_\mu\bar \psi}{\abs{F_S}^2}\right]-\frac{G\sigma^\mu\bar \psi}{2\abs{F_S}^2}\partial_\mu\left(\frac{\bar G^2}{2\bar F_S}\right)\\
	&+\frac{i}{2\bar F_S}\left(\frac{G}{F_S}\right)^2\partial_\mu\left(\frac{\bar G^2}{2\bar F_S}\right)\partial^\mu A-i\frac{G^2}{4\abs{F_S}^4}\left(\bar G\bar \sigma^\mu\sigma^\nu\partial_\nu\bar G\right)\partial_\mu A\\
	&-\frac{i}{4\abs{F_S}^2}\left\{\frac{G^2}{2F_S}\left[\bar G\bar \sigma^\nu\sigma^\mu\partial_\mu\left(\frac{\bar G}{\bar F_S}\right)\right]\partial_\mu A+\frac{\bar G^2}{2\bar F_S}\left[G\sigma^\nu\bar \sigma^\mu\partial_\nu\left(\frac{G}{F_S}\right)\right]\partial_\mu A\right\}\\
	&-\frac{1}{8|F_S|^2}G\sigma^\mu\bar{G}\[F_\phi\,\partial_\mu\(\frac{G}{F_S}\)^2+\bar{F}_\phi\,\partial_\mu\(\frac{\bar{G}}{\bar{F}_S}\)^2\].
	\end{aligned} \label{eqn:gc_B4}
\end{align}

\noindent We don't write explicit the expressions for $B^{(6)}$ and $B^{(8)}$ since they are not useful for our discussion. 

Let us turn now to the constraint (\ref{eqn:gc_ferm}) imposed  alone. This removes the fermion $\psi$ as a function of the goldstino, the scalar $\phi$ and the auxiliary fields. The correspondent starting equation is
\begin{equation}
\begin{aligned}
	\psi_\alpha=&i\frac{\left(\sigma^\mu\bar{G}\right)_\alpha}{\bar{F}_S}\partial_\mu\phi+\frac{F_\phi}{F_S}G_\alpha+i\frac{G\sigma^\mu\partial_\mu\bar{G}}{\abs{F_S}^2}\psi_\alpha+\frac{\left(\sigma^\nu\bar{\sigma}^\mu G\right)_\alpha}{\abs{F_S}^2}\partial_\mu\left(\frac{\bar{G}^2}{2\bar{F}_S}\right)\partial_\nu\phi\\
	&-i\frac{\left(\sigma^\mu\bar{G}\right)_\alpha}{\bar{F}_S}\frac{G\partial_\mu\psi}{F_S}+\frac{i}{2}\left(\frac{G}{F_S}\right)^2\frac{\left(\sigma^\mu\partial_\mu\bar{G}\right)_\alpha}{\bar{F}_S} F_\phi+\frac{i}{2}\left(\frac{G}{F_S}\right)^2\frac{\left(\sigma^\mu\bar{G}\right)_\alpha}{\bar{F}_S}\partial_\mu F_\phi
\\
	&+\frac{\left(\sigma^\nu\bar{\sigma}^\mu\partial_\nu\psi\right)_\alpha}{\abs{F_S}^2}\frac{G^2}{2F_S}\partial_\mu\left(\frac{\bar{G}^2}{2\bar{F}_S}\right)+\frac{1}{\abs{F_S}^2}\frac{G^2}{2F_S}\Box\left(\frac{\bar{G}^2}{2\bar{F}_S}\right)\psi_\alpha \ .
\end{aligned}
\label{eqn:gc_eq_psi}
\end{equation}
Again, we can solve this equation iteratively via an expansion of operators with increasing number of fermions, obtaining
\begin{equation}
	\psi_\alpha=\psi_\alpha^{(1)}+\psi_\alpha^{(3)}+\psi_\alpha^{(5)}+\psi_\alpha^{(7)}	\ ,
\label{eqn:gc_psi_exp}
\end{equation}
where
\begin{align}
	&\psi_\alpha^{(1)}=\frac{F_\phi}{F_S}G_\alpha+i\frac{\left(\sigma^\mu\bar G\right)_\alpha}{\bar F_S}\partial_\mu\phi \ , 
\label{eqn:gc_psi_1}\\\notag\\
&\begin{aligned}
	\psi_\alpha^{(3)}=&\frac{\left(\sigma^\nu\bar \sigma^\mu G\right)_\alpha}{\abs{F_S}^2}\partial_\mu\left(\frac{\bar G^2}{2\bar F_S}\right)\partial_\nu\phi-\frac{i}{2}\left(\frac{G}{F_S}\right)^2\frac{\left(\sigma^\mu\bar G\right)_\alpha}{\bar F_S}\partial_\mu F_\phi-\frac{i}{2}\frac{\left(\sigma^\mu\bar G\right)_\alpha}{\bar F_S}\partial_\mu\left(\frac{G}{F_S}\right)^2 F_\phi\\
	&+\frac{\left(\sigma^\mu\bar G\right)_\alpha}{\abs{F_S}^2}\left[G\sigma^\nu\partial_\mu\left(\frac{\bar G}{\bar F_S}\right)\right]\partial_\nu\phi+\left(\frac{\bar G}{\bar F_S}\right)^2\frac{G_\alpha}{2F_S}\Box\phi-\frac{G\sigma^\mu\partial_\mu\bar G}{\abs{F_S}^2}\frac{\left(\sigma^\nu\bar G\right)_\alpha}{\bar F_S}\partial_\nu\phi \ .
\end{aligned} \label{eqn:gc_psi_3}
\end{align}

\noindent Again, we don't display $\psi^{(5)}_\alpha$ and $\psi^{(7)}_\alpha$ being them not relevant for the discussion.

\section{Derivation of (\ref{lagAfterIntPhia})}\label{appendix:IRlagNoF}

The effective lagrangian \eqref{lagAfterIntPhia} can be derived as follows. The K\"ahler and superpotential are expanded in a series in $Q_L^{(a)}$, which respectively truncates at second and first order due to the nilpotency of $Q_L^{(a)}$,
\beq
\bead
&K(S,\bar S,Q_L,\bar Q_L)=K(S,\bar S,Q'_L,\bar Q'_L)\\
&\qquad\qquad\qquad\qquad+\(Q_L^{(a)}\partial_{Q_L}K(0,\bar S,Q'_L,\bar Q'_L)+h.c.\)+Q_L^{(a)}\bar Q_L^{(a)}\partial^2_{Q_L,\bar Q_L}K(0,0,Q'_L,\bar Q'_L) \ ,\\
&W(S,Q_L)=W(S,Q'_L)+Q_L^{(a)}\partial_{Q_L}W(0,Q'_L) \ .
\eead
\eeq
The $S$-dependence of the arguments is affected by $SQ^{(a)}_L=0$, since the above functions can also be (finitely) Taylor-expanded in $S$. Then, one makes use of the explicit expression of $Q^{(a)_L}$ in \eqref{soleomPhia} and of $\int d^2\theta \bar D^2f=-4\int d^4\theta f$ under a spacetime integration to write
\small
\beq
\int d^2\theta Q_L^{(a)}\partial_{Q_L}W(0,Q'_L)=-4\int d^4\theta \frac{\abs{S}^2\[4\partial_{\bar Q_L}\bar W(0,\bar Q_L')-\partial^2_{S,\bar Q_L} K(0,0,Q_L',\bar Q_L')D^2S\]}{\partial^2_{Q_L,\bar Q_L} K(0,0, Q_L',\bar Q_L')\abs{D^2S}^2}\partial_{Q_L}W(0,Q'_L) \ . 
\eeq
\normalsize
In the linear term of the K\"ahler expansion, one integrates by parts ($\int d^4\theta D^2 fg=\int d^4\theta f D^2 g$) the $D^2$ which arises from the expression of $Q^{(a)_L}$ in \eqref{soleomPhia},
\small
\beq
\bead
Q_L^{(a)}\partial_{Q_L}K(0,\bar S,Q'_L,\bar Q'_L)\to&\frac{\abs{S}^2\[4\partial_{\bar Q_L}\bar W(0,\bar Q_L')-\partial^2_{S,\bar Q_L} K(0,0,Q_L',\bar Q_L')D^2S\]}{\partial^2_{Q_L,\bar Q_L} K(0,0, Q_L',\bar Q_L')\abs{D^2S}^2}\bar D^2\partial_{Q_L}K(0,\bar S,Q'_L,\bar Q'_L)\\
=&\,\frac{\abs{S}^2\[4\partial_{\bar Q_L}\bar W(0,\bar Q_L')-\partial^2_{S,\bar Q_L} K(0,0,Q_L',\bar Q_L')D^2S\]}{\partial^2_{Q_L,\bar Q_L} K(0,0, Q_L',\bar Q_L')\abs{D^2S}^2}\\
&\times\(\bar D^2\bar S\partial^2_{\bar S,Q_L}K(0,0,Q'_L,\bar Q'_L)+\[\bar D\bar Q'_L\]^2\partial^3_{Q_L,\bar Q_L^2}K(0,0, Q_L',\bar Q_L')\)\ ,
\eead
\eeq
\normalsize
where the last equality relies on the nilpotency of $S$ and on the constraint on $Q'_L$. The third line of \eqref{lagAfterIntPhia} corresponds to the second piece in the last parenthesis above. Twice the second line of \eqref{lagAfterIntPhia} is obtained by adding the first piece in the last parenthesis above to the superpotential contribution and to their hermitian conjugates. The final result is halved by the addition of
\beq
Q_L^{(a)}\bar Q_L^{(a)}\partial^2_{Q_L,\bar Q_L}K(0,0,Q'_L,\bar Q'_L)=\frac{\abs{4\partial_{Q_L}W(0,\bar Q_L')-\partial^2_{Q_L,\bar S} K(0,0, Q_L',\bar Q_L')\bar D^2\bar S}^2}{\partial^2_{Q_L,\bar Q_L}K(0,0, Q_L',\bar Q_L')}\abs{\frac{S}{\bar D^2\bar S}}^2 \ .
\eeq

\end{appendices} 
\newpage


\end{document}